\documentclass[
]{aa}  

\usepackage{txfonts}

\usepackage{amsmath}
\usepackage{graphicx}
\usepackage{xcolor}
\usepackage{xparse}
\usepackage{bm}
\usepackage{latexsym}
\usepackage[
capitalize,
noabbrev
]{cleveref}
\usepackage{tikz}
\usepackage[normalem]{ulem}
\usepackage{paralist}
\usepackage{multirow}

\newcommand{\Bb}{\ensuremath{\bm{b}}}

\newcommand{\Bv}{\ensuremath{\bm{v}}}

\newcommand{\Bz}{\ensuremath{\bm{z}}}

\NewDocumentCommand{\curl}{o o}{\ensuremath{\IfNoValueTF{#2}{\nabla}{\nabla_{#2}} \times #1}}
\NewDocumentCommand{\grad}{o o}{\ensuremath{\IfNoValueTF{#2}{\nabla}{\nabla_{#2}} #1}}
\NewDocumentCommand{\diverg}{o o}{\ensuremath{\IfNoValueTF{#2}{\nabla}{\nabla_{#2}} \cdot #1}} 

\NewDocumentCommand{\years}{o}{\ensuremath{
\IfNoValueF{#1}{#1 \,}
\mathrm{years}
}}

\NewDocumentCommand{\days}{o}{\ensuremath{
\IfNoValueF{#1}{#1 \,}
\mathrm{days}
}}

\NewDocumentCommand{\months}{o}{\ensuremath{
\IfNoValueF{#1}{#1 \,}
\mathrm{months}
}}

\NewDocumentCommand{\s}{o}{\ensuremath{
\IfNoValueF{#1}{#1 \,}
\mathrm{s}
}}

\NewDocumentCommand{\nT}{o}{\ensuremath{
\IfNoValueF{#1}{#1 \,}
\mathrm{nT}
}}

\NewDocumentCommand{\km}{o}{\ensuremath{
\IfNoValueF{#1}{#1 \,}
\mathrm{km}
}}

\NewDocumentCommand{\au}{o}{\ensuremath{
\IfNoValueF{#1}{#1 \,}
\mathrm{AU}
}}

\NewDocumentCommand{\amu}{o}{\ensuremath{
\IfNoValueF{#1}{#1 \,}
\mathrm{amu}
}}

\NewDocumentCommand{\temp}{o}{\ensuremath{
\IfNoValueF{#1}{#1 \times}
\mathrm{10^5 \; K}
}}

\NewDocumentCommand{\cc}{o}{\ensuremath{
\IfNoValueF{#1}{#1 \;}
\mathrm{cm}^{-3}
}}

\NewDocumentCommand{\pct}{o}{\ensuremath{
\IfNoValueF{#1}{#1 \;}
\%
}}

\NewDocumentCommand{\Rs}{o}{\ensuremath{
\IfNoValueF{#1}{#1 \;}
\mathrm{R_S}
}}

\NewDocumentCommand{\kms}{o}{\ensuremath{
\IfNoValueF{#1}{#1 \;}
\mathrm{km \, s^{-1}}
}}

\NewDocumentCommand{\mWcc}{o}{\ensuremath{
\IfNoValueF{#1}{#1 \;}
\mathrm{mW \cc}
}}

\NewDocumentCommand{\eV}{o}{\ensuremath{
\IfNoValueF{#1}{#1 \;}
\mathrm{eV}
}}

\NewDocumentCommand{\keV}{o}{\ensuremath{
\IfNoValueF{#1}{#1 \;}
\mathrm{keV}
}}

\NewDocumentCommand{\MeV}{o}{\ensuremath{
\IfNoValueF{#1}{#1 \;}
\mathrm{MeV}
}}

\NewDocumentCommand{\nucleon}{s o}{\ensuremath{
\IfNoValueF{#2}{#2 \;}
\IfBooleanTF{#1}{\mathrm{nucleon}}{\mathrm{nuc}}
}}

\NewDocumentCommand{\MeVnuc}{s o}{\ensuremath{
\IfNoValueF{#2}{#2 \;}
\MeV \! /\IfBooleanTF{#1}{\nucleon*}{\nucleon}
}}

\NewDocumentCommand{\keVe}{o}{\ensuremath{
\IfNoValueF{#1}{#1 \;}
\mathrm{keV/e}
}}

\NewDocumentCommand{\n}{o}{\ensuremath{n
\IfNoValueF{#1}{_{#1}}
}}

\NewDocumentCommand{\Element}{m}{\ensuremath{\mathrm{#1}}}
\NewDocumentCommand{\QState}{m m}{\ensuremath{\mathrm{#1}^{#2+}}}

\NewDocumentCommand{\Hy}{o}{\IfNoValueTF{#1}{\Element{H}}{\QState{H}{#1}}}
\NewDocumentCommand{\He}{o}{\IfNoValueTF{#1}{\Element{He}}{\QState{He}{#1}}}
\NewDocumentCommand{\C}{o}{\IfNoValueTF{#1}{\Element{C}}{\QState{C}{#1}}}
\NewDocumentCommand{\N}{o}{\IfNoValueTF{#1}{\Element{N}}{\QState{N}{#1}}}
\NewDocumentCommand{\Ox}{o}{\IfNoValueTF{#1}{\Element{O}}{\QState{O}{#1}}}
\NewDocumentCommand{\Ne}{o}{\IfNoValueTF{#1}{\Element{Ne}}{\QState{Ne}{#1}}}
\NewDocumentCommand{\Mg}{o}{\IfNoValueTF{#1}{\Element{Mg}}{\QState{Mg}{#1}}}
\NewDocumentCommand{\Si}{o}{\IfNoValueTF{#1}{\Element{Si}}{\QState{Si}{#1}}}
\NewDocumentCommand{\Su}{o}{\IfNoValueTF{#1}{\Element{S}}{\QState{S}{#1}}}
\NewDocumentCommand{\Ca}{o}{\IfNoValueTF{#1}{\Element{Ca}}{\QState{Ca}{#1}}}
\NewDocumentCommand{\Fe}{o}{\IfNoValueTF{#1}{\Element{Fe}}{\QState{Fe}{#1}}}

\NewDocumentCommand{\FIP}{o O{=}}{\ensuremath{
\ensuremath{\mathrm{FIP}}\IfNoValueF{#1}{#2\eV[#1]}
}}

\NewDocumentCommand{\Qratio}{m m m}{\ensuremath{#1[#2] / #1[#3]}}

\NewDocumentCommand{\Qavg}{m o}{\ensuremath{\langle Q_{#1} \rangle \IfNoValueF{#2}{= #2}
}}

\NewDocumentCommand{\AbSEP}{s O{X} O{\Ox}}{\IfBooleanTF{#1}{\ensuremath{(#2/#3)}:\ensuremath{(#2/#3)_\mathrm{photo}}}
{\ensuremath{#2/#3}}}
\NewDocumentCommand{\AbSW}{s O{X} O{\Hy}}{\IfBooleanTF{#1}{\ensuremath{(#2/#3)}:\ensuremath{(#2/#3)_\mathrm{photo}}}
{\ensuremath{#2/#3}}}

\NewDocumentCommand{\PLawExp}{s o}{\ensuremath{b
\IfNoValueF{#2}{\IfBooleanTF{#1}{\approx}{=} #2}}}

\newcommand{\he}{\Element{He}}

\NewDocumentCommand{\M}{o}{\ensuremath{
\IfNoValueTF{#1}{\mathrm{M}}{(\mathrm{M})_{#1}}}}

\NewDocumentCommand{\Q}{s o}{\ensuremath{
\IfNoValueTF{#2}{\mathrm{Q}}{
\IfBooleanTF{#1}{\langle \mathrm{Q}_{#2} \rangle}{\mathrm{Q}_{#2}}
}}}

\NewDocumentCommand{\MpQ}{o}{\ensuremath{
\IfNoValueTF{#1}{\mathrm{M/Q}}{(\mathrm{M/Q})_{#1}}}}

\NewDocumentCommand{\As}{s o O{=}}{\ensuremath{A_s
\IfNoValueF{#2}{
#3 #2
\IfBooleanF{#1}{\%}}
}}

\NewDocumentCommand{\vs}{s o O{=} d{_}{_}}{\ensuremath{v_{s\IfNoValueF{#4}{;#4}}
\IfNoValueF{#2}{
\IfNoValueTF{#3}{=}{#3}
\IfBooleanTF{#1}{#2}{\kms[#2]}}
}}

\NewDocumentCommand{\vel}{s o O{=} d{_}{_}}{\ensuremath{v
\IfNoValueF{#4}{_{#4}}
\IfNoValueF{#2}{
\IfNoValueTF{#3}{=}{#3}
\IfBooleanTF{#1}{#2}{\kms[#2]}}
}}

\NewDocumentCommand{\vsw}{s o O{=}}{\ensuremath{v_\sw
\IfNoValueF{#2}{
\IfNoValueTF{#3}{=}{#3}
\IfBooleanTF{#1}{#2}{\kms[#2]}}
}}

\NewDocumentCommand{\vv}{s o O{=}}{\ensuremath{v_v
\IfNoValueF{#2}{
\IfNoValueTF{#3}{=}{#3}
\IfBooleanTF{#1}{#2}{\kms[#2]}}
}}

\NewDocumentCommand{\grate}{o o}{\ensuremath{
\gamma\IfNoValueF{#1}{/\Omega_{#1}}
\IfNoValueF{#2}{= 10^{{#2}}}
}}

\NewDocumentCommand{\gmax}{o}{\ensuremath{
\gamma_\mathrm{max}\IfNoValueF{#1}{/\Omega_{#1}}
}}

\NewDocumentCommand{\kvec}{o}{\ensuremath{
\vec{k} \rho\IfNoValueF{#1}{{_{#1}}}
}}

\NewDocumentCommand{\kpar}{o}{\ensuremath{
{k_\parallel} \rho\IfNoValueF{#1}{{_{#1}}}
}}

\NewDocumentCommand{\kper}{o}{\ensuremath{
{k_\perp} \rho\IfNoValueF{#1}{{_{#1}}}
}}

\NewDocumentCommand{\ani}{s o}{\ensuremath{
R\IfNoValueF{#2}{_{#2}}
\IfBooleanT{#1}{\, [\perp\!/\!\parallel]}
}}

\NewDocumentCommand{\Temp}{o}{\ensuremath{T{\IfNoValueF{#1}{_{#1}}}}}

\NewDocumentCommand{\Trat}{s m m o}{\ensuremath{
T_{\IfNoValueF{#4}{{#4};}#2}/T_{\IfNoValueF{#4}{{#4};}#3}
 \IfBooleanT{#1}{\, [\#]}
}}

\NewDocumentCommand{\pbeta}{s o}{\ensuremath{
\beta\IfNoValueF{#2}{_{#2}}
 \IfBooleanT{#1}{\, [\#]}
}}

\NewDocumentCommand{\pbetaR}{o}{\ensuremath{
(\pbeta[\parallel
\IfNoValueF{#1}{;#1}], \ani[#1])
}}

\NewDocumentCommand{\dv}{o}{\ensuremath{\Delta v\IfNoValueF{#1}{_{#1}}}}
\NewDocumentCommand{\ca}{o}{\ensuremath{C_{A\IfNoValueF{#1}{;#1}}}}
\NewDocumentCommand{\dvca}{o o}{\ensuremath{\dv[#1]/\ca[#2]}}

\NewDocumentCommand{\nuc}{o}{\ensuremath{\nu_{c\IfNoValueF{#1}{;#1}}}}
\NewDocumentCommand{\Nc}{o}{\ensuremath{N_{c\IfNoValueF{#1}{;#1}}}}
\NewDocumentCommand{\Ac}{o}{\ensuremath{A_{c\IfNoValueF{#1}{;#1}}}}

\NewDocumentCommand{\tauEXP}{o}{\ensuremath{
\tau_{\mathrm{exp}\IfNoValueF{#1}{;#1}
}}}

\NewDocumentCommand{\tauCC}{o}{\ensuremath{
\tau_{\mathrm{C}\IfNoValueF{#1}{;#1}
}}}

\NewDocumentCommand{\SSN}{o}{\ensuremath{\mathrm{SSN}
\IfNoValueF{#1}{#1}}}
\NewDocumentCommand{\NSSN}{o}{\ensuremath{\mathrm{NSSN}
\IfNoValueF{#1}{#1}}}

\newcommand{\sw}{\ensuremath{\mathrm{sw}}}

\NewDocumentCommand{\qpar}{o}{\ensuremath{
q_{\parallel
\IfNoValueF{#1}{;#1}
}}}

\NewDocumentCommand{\edv}{o}{\ensuremath{
\tilde{E}_{\dv[#1]
}}}

\NewDocumentCommand{\ndays}{o}{
\ensuremath{N_\mathrm{days}{\IfNoValueF{#1}{= {#1}}}}
}

\NewDocumentCommand{\se}{o}{\ensuremath{
S{\IfNoValueF{#1}{_{#1}}}
}}

\NewDocumentCommand{\ab}{o}{\ensuremath{
A{\IfNoValueF{#1}{_{#1}}}
}}

\NewDocumentCommand{\ahe}{o O{=}}{\ensuremath{\ab[\he]
\IfNoValueF{#1}{
\IfNoValueTF{#2}{=}{#2}#1\%}
}}

\NewDocumentCommand{\corr}{o}{\ensuremath{
\rho
\IfNoValueF{#1}{(#1)}
}}

\NewDocumentCommand{\xhel}{o O{=} o}{\ensuremath{\sigma_{c
\IfNoValueF{#3}{,#3}}
\IfNoValueF{#1}{
\IfNoValueTF{#2}{=}{#2}
#1}
}}

\NewDocumentCommand{\SpecInd}{o}{\ensuremath{\gamma
\IfNoValueF{#1}{_{#1}}}}
\NewDocumentCommand{\QT}{o}{\ensuremath{\mathrm{QT}
\IfNoValueF{#1}{= #1}}}

\NewDocumentCommand{\rsq}{o O{=}}{\ensuremath{R^2
\IfNoValueF{#1}{
\IfNoValueTF{#2}{=}{#2}
#1}
}}

\NewDocumentCommand{\func}{m m o O{=}}{\ensuremath{#1\left(#2\right)\IfNoValueF{#3}{#4 #3}}
}

\newcommand{\citepossessive}[1]{\citeauthor{#1}'s (\citeyear{#1})}

\definecolor{C0}{HTML}{1f77b4}
\definecolor{C1}{HTML}{ff7f0e}
\definecolor{C2}{HTML}{2ca02c}
\definecolor{C3}{HTML}{d62728}
\definecolor{C4}{HTML}{9467bd}
\definecolor{C5}{HTML}{8c564b}

\definecolor{QTFitGreen}{HTML}{2ca02c}
\definecolor{DodgerBlue}{HTML}{1e90ff}
\definecolor{Fuchsia}{HTML}{ff00ff}
\definecolor{TabGreen}{HTML}{2ca02c}
\definecolor{Cyan}{HTML}{00ffff}
\definecolor{LimeGreen}{HTML}{32cd32}
\definecolor{Lime}{HTML}{00ff00}

\definecolor{MaxPink}{HTML}{e377c2}
\definecolor{MinPurple}{HTML}{9467bd}

\definecolor{q}{HTML}{228B22}
\definecolor{wc}{HTML}{FF8C00}
\definecolor{dnc}{HTML}{FF00FF}
\definecolor{todo}{HTML}{e13748}
\definecolor{ben}{HTML}{e13748}
\definecolor{bob}{HTML}{0080FF}

\NewDocumentCommand{\answer}{s o m}{\IfBooleanF{#1}{\textcolor{q}{\textbf{A}\IfNoValueF{#2}{ (#2)}: \textit{#3}}}}

\NewDocumentCommand{\wc}{s m}{\IfBooleanTF{#1}{#2}{\textcolor{wc}{\textbf{WC:} \textit{#2}}}}
\NewDocumentCommand{\ws}{s m}{\IfBooleanTF{#1}{#2}{\textcolor{wc}{\textbf{WS:} \textit{#2}}}}

\NewDocumentCommand{\delete}{s m}{\IfBooleanF{#1}{\textcolor{todo}{\textbf{Delete:} \textit{#2}}}}
\NewDocumentCommand{\todo}{s o m}{\IfBooleanF{#1}{\textcolor{todo}{\textbf{TODO}\IfNoValueF{#2}{ (#2)}: \textit{#3}}}}
\NewDocumentCommand{\verify}{s o m}{\IfBooleanTF{#1}{#3}{\textcolor{todo}{\textbf{VERIFY}\IfNoValueF{#2}{ (#2)}: \textit{#3}}}}
\NewDocumentCommand{\goal}{s o m}{\IfBooleanTF{#1}{#3}{\textcolor{todo}{\textbf{GOAL}\IfNoValueF{#2}{ (#2)}: \textit{#3}}}}

\NewDocumentCommand{\move}{s o m}{\textcolor{dnc}{\textbf{\IfBooleanTF{#1}{Duplicate}{Move}}\IfNoValueF{#2}{ (#2)}: \textit{#3}}}
\NewDocumentCommand{\dupe}{o m}{\move*[#1]{#2}}
\NewDocumentCommand{\intro}{s m}{\IfBooleanTF{#1}{\dupe[Intro]{#2}}{\move[Intro]{#2}}}
\NewDocumentCommand{\dnc}{s m}{\IfBooleanTF{#1}{\dupe[DnC]{#2}}{\move[DnC]{#2}}}
\NewDocumentCommand{\fw}{s m}{\IfBooleanTF{#1}{\dupe[Future Work]{#2}}{\move[DnC]{#2}}}

\DeclareDocumentCommand{\EmptyTimes}{O{black}}{\ensuremath{\mathord{\begin{tikzpicture}[line width=0.2ex, x=1.5ex, y=1.5ex]
\draw[color=#1] (0, 0.25) -- (0.25, 0.5) -- (0, 0.75) -- (0.25, 1.0) -- (0.5, 0.75) -- (0.75, 1.0) -- (1.0, 0.75) -- (0.75, 0.5) -- (1.0, 0.25) -- (0.75, 0) -- (0.5, 0.25) -- (0.25, 0) -- cycle;
\end{tikzpicture}}}}

\DeclareDocumentCommand{\SolidBand}{O{black} D{<}{>}{1}}{\ensuremath{\mathord{\begin{tikzpicture}[line width=1.25ex, x=1.25ex, y=1.25ex, yshift=5ex]
\draw[color=#1, opacity=#2] (0,0.5) -- (1.5,0.5);
\draw[opacity=0, line width=0.1ex] (0,0) -- (1.5,0);
\end{tikzpicture}}}}

\DeclareDocumentCommand{\SolidBandVertLines}{O{black} D{<}{>}{1}}{\ensuremath{\mathord{\begin{tikzpicture}[line width=1.25ex, x=1.25ex, y=1.25ex, yshift=5ex]
\draw[color=#1, opacity=#2] (0,0.5) -- (1.5,0.5);
\draw[opacity=0, line width=0.1ex] (0,0) -- (1.5,0);
\draw[color=#1, line width=0.2ex] (0, 0) -- (0, 1);
\draw[color=#1, line width=0.2ex] (1.5, 0) -- (1.5, 1);
\end{tikzpicture}}}}

\DeclareDocumentCommand{\SolidLine}{O{black}}{\ensuremath{\mathord{\begin{tikzpicture}[line width=0.3ex, x=1.25ex, y=1.25ex, yshift=5ex]
\draw[color=#1] (0,0.5) -- (1,0.5);
\draw[opacity=0] (0,0) -- (1,0);
\end{tikzpicture}}}}

\DeclareDocumentCommand{\DashedLine}{O{black} o D{<}{>}{1}}{\ensuremath{\mathord{\begin{tikzpicture}[line width=0.3ex, x=1.25ex, y=1.25ex, yshift=5ex]
\IfNoValueF{#2}{\draw[color=#2, opacity=#3] (0, 0.5) -- (1.75, 0.5);}
\draw[color=#1, opacity=#3] (0,0.5) -- (0.75,0.5);
\draw[color=#1, opacity=#3] (1.0,0.5) -- (1.75,0.5);
\draw[opacity=0] (0,0) -- (1.25,0);
\end{tikzpicture}}}}

\DeclareDocumentCommand{\HighlightDashedLine}{O{black} O{LimeGreen} D{<}{>}{1}}{\ensuremath{\mathord{\begin{tikzpicture}[line width=0.3ex, x=1.25ex, y=1.25ex, yshift=5ex]
\draw[color=#2, opacity=#3, line width=0.8ex] (0, 0.5) -- (1.75, 0.5);
\draw[color=#1] (0,0.5) -- (0.75,0.5);
\draw[color=#1] (1.0,0.5) -- (1.75,0.5);
\draw[opacity=0] (0,0) -- (1.25,0);
\end{tikzpicture}}}}

\DeclareDocumentCommand{\DotDotDotLine}{O{white} O{black}}{\ensuremath{\mathord{\begin{tikzpicture}[line width=0.3ex, x=1.25ex, y=1.25ex, yshift=5ex]
\draw[color=#1] (0, 0.5) -- (1.5, 0.5);
\draw[color=#2] (0,0.5) -- (0.3,0.5);
\draw [color=#2](0.6,0.5) -- (0.9,0.5);
\draw[color=#2] (1.2,0.5) -- (1.5,0.5);
\draw[opacity=0] (0,0) -- (1.5,0);
\end{tikzpicture}}}}

\DeclareDocumentCommand{\DotDotDotLineTwo}{O{yellow} O{black}}{\ensuremath{\mathord{\begin{tikzpicture}[line width=0.3ex, x=1.25ex, y=1.25ex, yshift=5ex]
\draw[color=#1] (0,0.5) -- (1.5,0.5);
\draw[color=#2] (0,0.5) -- (0.3,0.5);
\draw[color=#2] (0.6,0.5) -- (0.9,0.5);
\draw[color=#2] (1.2,0.5) -- (1.5,0.5);
\draw[opacity=0] (0,0) -- (1.5,0);
\end{tikzpicture}}}}

\DeclareDocumentCommand{\DashDotDotDotLine}{O{white} O{black}}{\ensuremath{\mathord{\begin{tikzpicture}[line width=0.3ex, x=1.25ex, y=1.25ex, yshift=5ex]
\draw[color=#1] (0, 0.5) -- (2.8, 0.5);
\draw[color=#2] (0,0.5) -- (1,0.5);
\draw[color=#2] (1.3,0.5) -- (1.6,0.5);
\draw[color=#2] (1.9,0.5) -- (2.2,0.5);
\draw[color=#2] (2.5,0.5) -- (2.8,0.5);
\draw[opacity=0] (0,0) -- (2.8,0);
\end{tikzpicture}}}}

\DeclareDocumentCommand{\Circle}{O{black}}{\ensuremath{\mathord{\begin{tikzpicture}[line width=0.3ex, x=1.25ex, y=1.25ex, yshift=5ex]
\draw[color=#1] circle (0.75ex);
\end{tikzpicture}}}}

\NewDocumentCommand{\sect}{o m}{Section~\ref{sec:#2}\IfNoValueF{#1}{ #1}}
\NewDocumentCommand{\eq}{o m}{\cref{eq:#2}\IfNoValueF{#1}{ #1}}
\NewDocumentCommand{\tbl}{o m}{\cref{tbl:#2}\IfNoValueF{#1}{ #1}}

  \usepackage{multirow}
\usepackage{tablefootnote}

\usepackage[switch]{lineno}

\NewDocumentCommand{\plotSWEOverTime}{s}{
\IfBooleanTF{#1}{\begin{figure*}}{\begin{figure}}
\begin{centering}
\includegraphics[width=\linewidth]{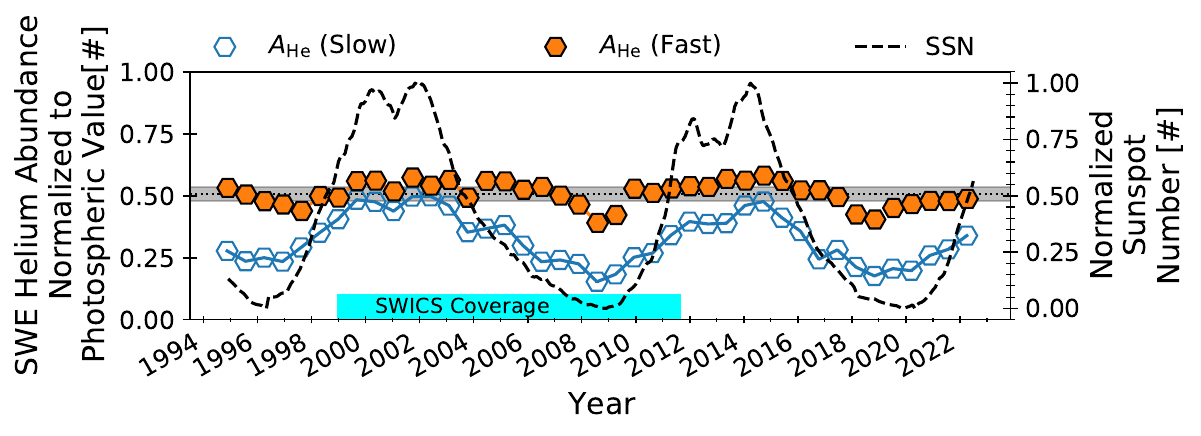}
\caption{\label{fig:t-swe}
The SWE helium abundance in fast (filled marker) and slow (empty marker) wind intervals as a function of time, aggregated down to 250 days.
The right axis plots the normalized 13-month smoothed sunspot number.
The horizontal dotted line surrounded by the gray band indicating \ahe[51 \pm 3] of its photospheric value is the weighted mean of the 250-day binned values along with the weighted uncertainty, where the weights are calculated as the standard deviation of \ahe\ in each 250-day bin.
}
\end{centering}
\IfBooleanTF{#1}{\end{figure*}}{\end{figure}}
}

\NewDocumentCommand{\plotSWICSOverTime}{s}{
\IfBooleanTF{#1}{\begin{figure*}}{\begin{figure}}
\begin{centering}
\includegraphics[width=\linewidth]{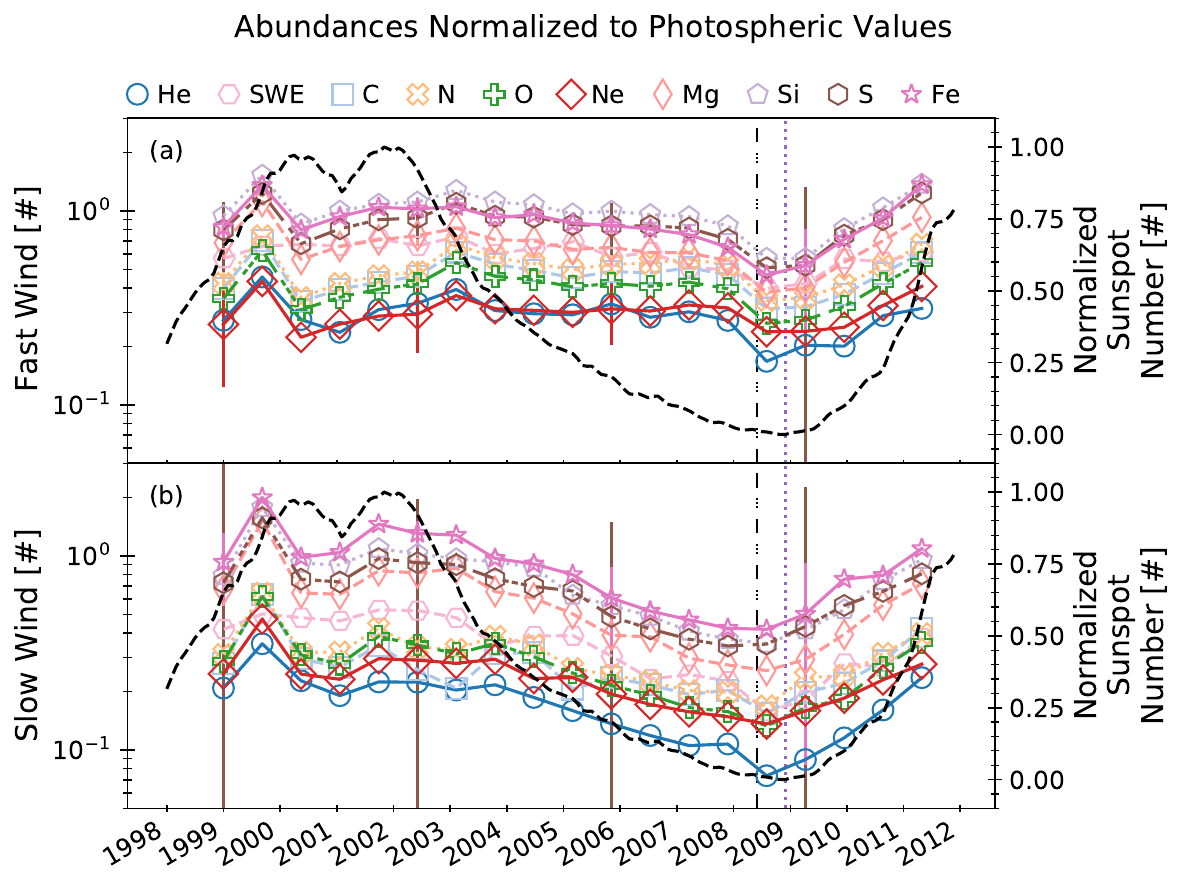}
\caption{\label{fig:t-swics}
The SWICS abundances normalized to their photospheric values in (a) fast and (b) slow wind as a function of time, aggregated in 250-day intervals.
The right axes plots the normalized 13-month smoothed sunspot number.
The SWE abundances normalized to its photospheric values are plotted for reference.
The vertical dotted lines indicate solar minimum 24.
The vertical dash-dotted line indicates the helium shutoff \citep{Alterman2021}.
}
\end{centering}
\IfBooleanTF{#1}{\end{figure*}}{\end{figure}}
}

\NewDocumentCommand{\plotChemistryFinal}{s}{
\IfBooleanTF{#1}{\begin{figure*}}{\begin{figure}}
\begin{centering}
\includegraphics[page=2, width=\linewidth]{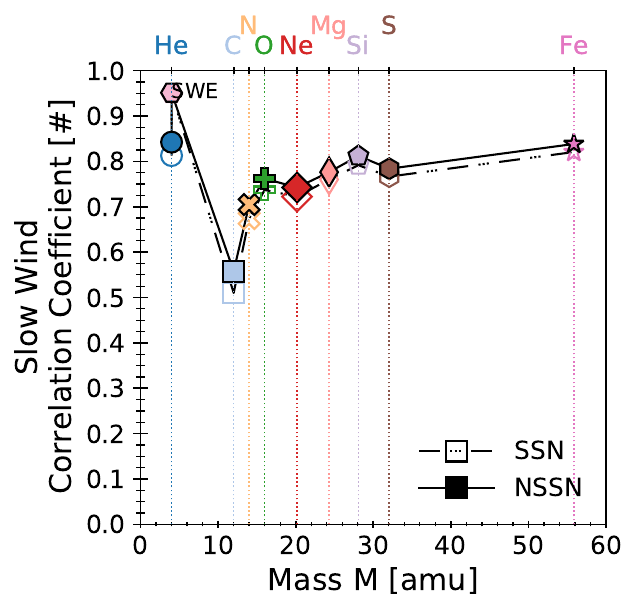}
\caption{\label{fig:chemistry-delay}
Correlation coefficient between \NSSN\ (solid line) or \SSN\ (dash-dotted line)  \AbSW* observed by ACE/SWICS along with \ahe\ observed by Wind/SWE.
With the exclusion of the helium abundances, the correlation coefficients monotonically increase with increasing \M.
All correlation coefficients using \NSSN\ are stronger than those using \SSN.
}
\end{centering}
\IfBooleanTF{#1}{\end{figure*}}{\end{figure}}
}

 \clearpage{}

\clearpage{}
\clearpage{}

\clearpage{}
\clearpage{}

\NewDocumentCommand{\TblCorr}{s}{
\IfBooleanTF{#1}{\begin{table*}}{\begin{table}}
\centering
\begin{tabular}{c cc cc}
\hline
{} & \multicolumn{2}{c}{Slow Wind} & \multicolumn{2}{c}{Fast Wind} \\
{} & SWE & \SSN & SWE & \SSN \\
\hline
SWE* & --- &   0.94 & --- &   0.66\\
SWE & --- &   0.92 & --- &   0.76\\
He & 0.83 &   0.81 & 0.54 &   0.49\\
C  & 0.47 &   0.49 &  --- &   ---\\
N  & 0.66 &   0.62 &  --- &   ---\\
O  & 0.78 &   0.72 & --- &   ---\\
Ne & 0.85 &   0.71 &  --- &   ---\\
Mg & 0.87 &   0.75 & 0.46 &   0.53\\
Si & 0.79 &   0.77 & 0.52 &   0.51\\
S  & 0.86 &   0.76 &  --- &   ---\\
Fe & 0.91 &   0.81 & 0.58 &  0.59\\
\hline
\end{tabular}
\caption{\label{tbl:corr}
Fast and slow wind correlation coefficients for the species identified in the first column .
The columns labeled \emph{SWE} provide the correlation coefficient between a given species and \ahe\ observed by the SWE Faraday cups.
Columns labeled \emph{SSN} give the correlation coefficient between a given abundance as \SSN.
The row labels \emph{SWE*} provides the correlation coefficients between \ahe\ and \SSN\ over the full time period plotted in \cref{fig:t-swe}.
The row labeled \emph{SWE} provides the same correlation coefficients calculated over the time period when SWICS data is available.
The correlation coefficients between \ahe\ observed by SWE and \SSN\ cover the full time period plotted in \cref{fig:t-swe}.
Only significant coefficients with p-values $< 0.05$ are shown.
}
\IfBooleanTF{#1}{\end{table*}}{\end{table}}
}

\RenewDocumentCommand{\TblCorr}{s}{
\IfBooleanTF{#1}{\begin{table*}}{\begin{table}}
\centering
\begin{tabular}{ccccccc}
\hline
{} & \multicolumn{3}{c}{Slow Wind} & \multicolumn{3}{c}{Fast Wind} \\
{} &  SWE &  SSN & NSSN &  SWE &  SSN & NSSN \\
\hline
SWE* &  --- & 0.94 & 0.95 &  --- & 0.66 & 0.72 \\
SWE &  --- & 0.95 & 0.95 &  --- & 0.75 & 0.76 \\
He  & 0.84 & 0.81 & 0.84 & 0.55 & 0.50 & 0.50 \\
C   & 0.62 & 0.51 & 0.56 &  --- &  --- &  --- \\
N   & 0.80 & 0.67 & 0.70 &  --- &  --- &  --- \\
O   & 0.84 & 0.74 & 0.76 &  --- &  --- &  --- \\
Ne  & 0.88 & 0.72 & 0.74 &  --- &  --- &  --- \\
Mg  & 0.90 & 0.76 & 0.78 & 0.58 & 0.53 & 0.59 \\
Si  & 0.90 & 0.79 & 0.81 & 0.62 & 0.53 & 0.58 \\
S   & 0.91 & 0.77 & 0.78 & 0.52 &  --- & 0.48 \\
Fe  & 0.93 & 0.82 & 0.84 & 0.72 & 0.59 & 0.66 \\
\hline
\end{tabular}
\caption{\label{tbl:corr}
Fast and slow wind correlation coefficients for the species identified in the first column .
The columns labeled \emph{SWE} provide the correlation coefficient between a given species and \ahe\ observed by the SWE Faraday cups.
Columns labeled \emph{SSN} and \emph{NSSN} give the correlation coefficient between a given abundance as \SSN\ and \NSSN, respectively.
The row labels \emph{SWE*} provides the correlation coefficients between \ahe\ and \SSN\ over the full time period plotted in \cref{fig:t-swe}.
The row labeled \emph{SWE} provides the same correlation coefficients calculated over the time period when SWICS data is available.
The correlation coefficients between \ahe\ observed by SWE and \SSN\ cover the full time period plotted in \cref{fig:t-swe}.
Only significant coefficients with p-values $< 0.05$ are shown.
}
\IfBooleanTF{#1}{\end{table*}}{\end{table}}
}
\clearpage{}

\graphicspath{{./}{figures/}}
\DeclareGraphicsExtensions{.pdf,.png,.jpg,.eps,}

\usepackage{lipsum}
\NewDocumentCommand{\BlindText}{O{6}}{\todo{Remove blind text. This is here to help figures render nicely.}
\textcolor{white}{\lipsum*[1-#1]}}

\begin{document}

\defcitealias{ACE:SWICS:FSTransition}{Composition Saturation Paper}
\defcitealias{Wind:SWE:ahe:xhel}{Helium Saturation Paper}

\newcommand{\satpoint}{\ensuremath{\left(\vs,\As\right)}}

\title{The Evolution of Heavy Ion Abundances with Solar Activity}

\newcommand{\Goddard}{Heliophysics Science Devision,
NASA Goddard Space Flight Center,
8800 Greenbelt, Road,
Greenbelt, MD, 20771, USA}

\newcommand{\CfA}{Center for Astrophysics $\vert$ Harvard \& Smithsonian,
\\60 Garden Street, Cambridge, MA 02138, USA}

\newcommand{\CLaSP}{University of Michigan \\
Department of Climate and Space Sciences \& Engineering \\
Climate and Space Research Building \\
2455 Hayward St.\\
Ann Arbor, MI, 48109, USA}

\newcommand{\IAPS}{
INAF - Institute for Space Astrophysics and Planetology\\
Via Fosso del Cavaliere, 100\\
00133 Rome, Italy
}

\author{B.\ L.\ Alterman\inst{1}
\and
Y.\ J.\ Rivera\inst{2}
\and
S.\ T.\ Lepri\inst{3}
\and
J.\ M.\ Raines \inst{3}
\and
R.\ D'Amicis \inst{4}
}

\institute{\Goddard\\
\email{b.l.alterman@nasa.gov}
\and
\CfA
\and
\CLaSP
\and
\IAPS
}

\date{Received September 15, 1996; accepted March 16, 1997}

\abstract{ When observed at \au[1], solar wind with slow speeds (\vsw[500][\lesssim]) is typically considered to have originated in source regions with magnetic topologies that are intermittently open to the heliosphere.
Solar wind with fast speeds (\vsw[500][\gtrsim]) is typically considered to have originated in source regions that are continuously open to the heliosphere, e.g coronal holes.
The evolution of the solar wind helium abundance (\ahe) with solar activity is likely driven by the evolution of different solar wind source regions.
Because slow wind is observed in the ecliptic more often than fast wind, a significant amount of the literature on this subject analyzes solar wind with slow and intermediate speeds \kms[\lesssim 600].
Utilizing the change in the gradient of \ahe\ with increasing \vsw, \citet{ACE:SWICS:FSTransition,Wind:SWE:ahe:xhel} have identified characteristic speeds at which our observations of the the abundance of solar wind helium and heavier elements transition from source regions that have intermittently to continuously open topologies.
}{ We aim to increase the maximum speed over which such analysis of the association between solar wind abundances and solar activity is performed to \kms[800], a rough upper limit on non-transient solar wind speeds when observed near \au[1].
We also aim to characterize the evolution of heavy element abundances \AbSW* with solar activity.
This analysis provides insight into the evolution of solar wind source regions with solar activity.
}{ We separate the solar wind into ``fast'' and ``slow'' for each element's abundance based on the characteristic speed derived for it in \citet{ACE:SWICS:FSTransition}.
We analyze the evolution of helium and heavy element abundances with solar activity using ACE/SWICS observations in each speed interval and correlate these abundances with solar activity as indicated by the 13-month smoothed sunspot number and a normalized version that accounts for the sunspot number's amplitude in each cycle.
Comparing the SWICS abundances with \ahe\ derived from Wind/SWE observations validates our analysis.
}{ We show that 
(1) \ahe\ strongly correlates with sunspot number in slow and fast wind; 
(2) the average non-transient solar wind \ahe\ is limited to $51\%$ of its photospheric value;
(3) slow wind heavy element abundances (with the exception of \C) do evolve significantly with solar activity;
(4) fast wind heavy element abundances do not significantly evolve with solar activity; 
(5) the correlation coefficient with sunspot number of elemental abundances for species heavier than \He\ monotonically increases with increasing mass; and 
(6) the correlation coefficients between the in situ observations and the normalized sunspot number are stronger than those using the unnormalized sunspot number.
We also report that the minimum in heavy element abundances may be closer to the rapid depletions and recoveries of \ahe\ that precede and predict sunspot minima, i.e.\ helium shutoff.
However higher time resolution analysis is necessary to properly characterize this signature.
}{ We infer that
(1) the sunspot number is a clock timing the solar cycle, but not the driver of the physical process underlying the evolution of \ahe\ and heavy element abundances with solar activity;
(2) this underlying process is likely related to the energy available to accelerate the solar plasma from the chromosphere and transition region or low corona into the solar wind; and
(3) the differences between the evolution of slow and fast solar wind \ahe\ and heavy element abundances are similarly related to the energy available to accelerate the elements at these heights above the Sun's surface.
}

   \maketitle

\section{Introduction \label{sec:intro}} 

Broadly, there are two classes of solar wind sources on the Sun.
Solar wind from sources with magnetic fields that are continuously open to interplanetary space like coronal holes \citep{Phillips1994,Geiss1995} tends to be accelerated to faster speeds and carry significant non-thermal features \citep{Kasper2008,Kasper2017,Tracy2015,Tracy2016,Wind:SWE:bimax,Stakhiv2016,Alterman2018,Berger2011,Klein2021,Verniero2020,Verniero2022,Durovcova2019}.
Sources with magnetic fields that are intermittently open to interplanetary space like helmet streamers, pseudostreamers, the boundaries between pseudostreamers and coronal holes (CHs), and the separatrix or S-Web \citep{Fisk1999,Subramanian2010,Antiochos2011,Crooker2012,Abbo2016,Antonucci2005} tend to accelerate slower wind that is more thermalized, i.e.\ the non-thermal features are smaller or non-existent.

Due to the difficulty in measuring absolute densities of heavy elements, solar wind abundances are often considered the ratio of the densities of two elements.
In this paper, we consider solar wind abundances of helium (\He) and heavier elements to the hydrogen (\Hy) number density ($\n[X]/\n[\Hy]$ for element $X$).
We denote such abundances as \AbSW\ and will often normalize them to their photospheric values \AbSW*.
We normalize them in this way because these abundances are set below the sonic critical point in the chromosphere and/or transition region  \citep{LR:FIP,Laming2004,Laming2009,Schwadron1999,Geiss1982a,Geiss1995b,Rakowski2012, Lepri2021,Rivera2022a}.
For the case of the helium abundance, we denote it as \ahe\ when it is not normalized to its photospheric value to maintain terminology consistent with prior work.
Slow and fast wind have distinct fractionation patterns that reflect differences in their source regions \citep{vonSteiger2000,Geiss1995,Geiss1995b,Zhao:InSituComposition:Sources,Zhao2022,Xu2014,Fu2017,Fu2015,Brooks2015,Weberg2015a,Zurbuchen2016,Ervin2024a,Rivera2025}.
As such, these abundances trace the source region features and can map \emph{in situ} observations back to their sources.
Several key results have been derived in this manner.

In slow and intermediate speed wind, the highly variable \ahe\ changes with solar activity \citep{Feldman1978}, which is typically quantified with the sunspot number (\SSN) \citep{SIDC}, the number of optically dark spots on the Sun's surface.
For solar wind speeds \vsw[500][\lesssim], the strength of the correlation between \ahe\ and \SSN\ is strongest in slower wind and decreases with increasing \vsw\ \citep{Aellig:Ahe,Kasper:Ahe,Alterman2019}.
Over this speed range, the modulation of \ahe\ is linear and \ahe\ drops below detectable levels at the vanishing speed \vv[259 \pm 12] \citep{Kasper:Ahe}.
The modulation of \ahe\ with solar activity also lags \SSN\ by 150 to 350 days, with a delay that increases with increasing \SSN, and is likely tied to changes in the magnetic topology of solar wind source regions on the Sun \citep{Alterman2019,Yogesh:Ahe}.

In contrast to the long duration evolution of \ahe\ with solar activity, \citet{Alterman2021} observed a rapid depletion and recovery of the helium abundance that is concurrent in time across all \vsw[601][\leq].
They refer to this as the ``helium shutoff''.
It precedes sunspot minima by 229 to 300 days and the uncertainty on this value is dominated by the 250-day averaging window.
They argue that this process is driven by a physical mechanism in or below the chromosphere and transition region because it impacts \ahe\ (which is set in these regions) irrespective of \vsw\ (which is set above these heights).
They further posit that the driving mechanism is related to the cancellation of the equatorial component of the buoyant toroidal magnetic flux near the equator that cancels during during the death of each solar cycle.

The abundances differentiate between solar wind from different types of source regions in a manner that is more rigorous than speed alone \citep{vonSteiger2000,Geiss1995,Geiss1995b,Zhao2022,Xu2014,Fu2017,Fu2015,ACE:SWICS:AUX}.
\citet{ACE:SWICS:AUX} show that heavy ion abundances referenced to \Hy\ are depleted in fast and slow wind during solar minima with respect to solar maxima, with the depletion being more significant in slow than fast wind.
The first ionization potential (\FIP) bias is the enhancement with respect to photospheric values of elements with \FIP[10][<] above high \FIP[10][>] elements.
Although both fast and slow abundances with respect to \Hy\ are depleted during solar minima in comparison to solar maxima, the fast wind's FIP bias does vary with solar activity and the slow wind's does not.
Synthesizing these different observations, the authors infer that average iron charge state (\Qavg{\Fe}) observations, ``suggest that the differences between fast and slow solar wind are mostly generated in the low corona, where \C\ and \Ox\ freeze-in, and then the two winds experience a similar evolution in the extended corona, where \Fe\ freezes-in \citep{ACE:SWICS:AUX}.''

Observations of quiescent streamers suggest that neither the coronal electron temperature nor density vary with solar activity and, as such, the observed changes in charge state ratios must be related to changes above the heights at which the quiescent streamers are observed or in the efficiency of solar wind acceleration over the solar cycle \citep{Landi2014}.
As different solar wind source regions likely accelerate solar wind with different efficiencies, this latter inference could also indicate a difference in the frequency at which solar wind originating in different source regions is observed over the solar cycle.
\citet{Kasper:Ahe} propose two sources of the slow wind.
\begin{compactenum}[1.]
\item Active regions (ARs), which produce solar wind with slow speeds and helium abundances that resemble fast wind, and are observed more often during solar maxima.
\item The streamer belt, which is highly variable, carries signatures of gravitational settling, and is primarily observed during solar minima.
\end{compactenum}
Models show that the coronal heat flux into the transition region modulates the solar wind speed by modifying coronal densities and electron heating in the transition region \citep{Lie-Svendsen2001,Lie-Svendsen2002}.
Comparisons between 
\begin{inparaenum}[(1)]
\item \ahe,
\item the solar wind iron-to-oxygen ratio normalized to its photospheric value (\AbSEP*[\Fe]), and
\item \Qavg{\Fe}
\end{inparaenum}
along with the variation of these quantities with both solar activity and the transition region (TR) scale size \citep{McIntosh:Ahe} suggest that the decrease in \ahe\ during solar minima is driven by the evolution of the coronal heat flux into the transition region with solar activity because changes in this heat flux impact the energy available the plasma at transition region depths.
This coupling is likely related to the local magnetic topology, which also implies that it varies with source region \citep[Section 4.3 and references therein]{Wind:SWE:ahe:xhel}, and therefore changes in the solar wind acceleration efficiency across these regions.

\plotSWEOverTime*
To statistically quantify the relationship between composition and changes in \vsw\ for source regions with continuously and intermittently open magnetic fields, \citet[hereafter \citetalias{Wind:SWE:ahe:xhel}]{Wind:SWE:ahe:xhel} have characterized the change in gradient of \ahe\ derived from the Wind Faraday cups \citep{Wind:SWE,Wind:SWE:bimax} as a function of \vsw\ (\grad[\ahe][\vsw]) with a bi-linear fit.
Here, \vsw\ is calculated as the bulk proton speed.
Because \grad[\ahe][\vsw] for speeds $v > \vs$ is approximately zero, they call this the helium saturation point \satpoint\ with saturation speed \vs\ and saturation abundance \As.
The saturation point is $\left( \vsw, \ahe \right) = \satpoint$.
With this technique, \citeauthor{Wind:SWE:ahe:xhel} have derived the saturation speed (\vs[433]) and abundance (\As[4.19]) at which \grad[\ahe][\vsw] changes from highly variable when \vsw*[\vs][<] (slow wind) to fixed at \As[4.19][\approx] for speeds \vsw*[\vs][>] (fast wind).

The normalized cross helicity \xhel\ is a measure of the correlation between fluctuations in the components of the solar wind's velocity and magnetic field.
It is given by $\xhel = \frac{e^+ - e^-}{e^+ + e^-}$, where $e^\pm = \frac{1}{2}\langle \left(z^\pm\right)^2\rangle$ are the energies in the Elsässer variables \citep{ElsasserVariables,Tu1989,Grappin1991}.
The Elsässer variables are $\Bz^\pm = \Bv \pm \frac{1}{\sqrt{\mu_0\rho}}\Bb$ for velocity \Bv, magnetic field \Bb, mass density $\rho$, and permeability of free space $\mu_0$.
The cross helicity can be used to quantify the how Alfvénic solar wind fluctuations are \citep{Tu1995,LR:turbulence,Woodham2018,Wind:SWE:ahe:xhel}.

Quantifying how \grad[\ahe][\vsw] and $\left(\vs,\As\right)$ change with the normalized cross helicity (\xhel), the \citetalias{Wind:SWE:ahe:xhel} argues that solar wind with speeds \vsw*[\vs][<] observed at \au[1] is predominantly from intermittently open source regions, while solar wind with speeds \vsw*[\vs][>] is predominantly from continuously open sources.
They draw this inference, in part, because \ahe\ is set in the chromosphere/transition region (below the sonic critical point), while \xhel\ is modified by motion in the corona (e.g. interchange reconnection) until the solar wind cross the Alfvén critical surface, above which \xhel\ primarily decays during propagation through interplanetary space.
Figure 2 from the \citetalias{Wind:SWE:ahe:xhel} is a cartoon illustrating how \ahe\ and \xhel\ are set at different heights above the solar surface along with the relationship between these quantities and the solar wind speed.

The Alfvénic slow wind (ASW) is solar wind with characteristically slow speeds but other properties that reflect typical fast wind properties including helium-to-hydrogen temperature ratios, non-zero \He\ and heavy ion flow velocities in excess of \Hy, and high levels of correlation between the \Hy\ velocity and magnetic field \citep{DAmicis2021a,DAmicis2021,DAmicis2018,Damicis2016,DAmicis2015,Yardley2024,Marsch1981,DAmicis2011a,Ervin2024,Rivera2025}.
 The \citetalias{Wind:SWE:ahe:xhel} characterizes \vsw\ as a function of \xhel\ and \ahe, revealing that ASW is an exception to the classification of solar wind with speeds slower and faster than \vs\ as from intermittently and continuously open source regions.
This is because ASW is the slow speed extension of solar wind born in continuously open regions.

To characterize the dependence of \satpoint\ on solar wind composition, 
\citet[hereafter \citetalias{ACE:SWICS:FSTransition}]{ACE:SWICS:FSTransition} applies the same bilinear or saturation fitting technique from the \citetalias{Wind:SWE:ahe:xhel} to heavy ion abundances \AbSW\ derived from Advanced Composition Explorer \citep[ACE,][]{ACE} Solar Wind Ion Composition Experiment \citep[SWICS,][]{ACE:SWICS} observations.
We use the 12-minute \Hy\ densities reported by SWICS,down-sampled to the heavy ion measurement cadence.
Because they are utilizing heavy ion composition observations, \citetalias{ACE:SWICS:FSTransition} normalize their abundances to their photospheric values \AbSW*.
As in \citetalias{ACE:SWICS:FSTransition}, we label the SWE helium abundance as \ahe\ and the SWICS helium abundance as \AbSW*[\He].
The \citetalias{ACE:SWICS:FSTransition} shows agreement between the helium saturation point derived from both Wind and ACE observations.
In the worst case, \vs\ is \kms[15] different between the two instruments and \As\ differs by at most 0.019 percentage points between the two instruments, indicating the reliability of extending the analysis of the \citetalias{Wind:SWE:ahe:xhel} from Wind/FC observations to ACE/SWICS observations.
They then show that the gradient of heavy ion abundances as a function of \vsw\ is independent of M, Q, M/Q, FIP, etc. in the slow wind, suggesting that the processes that set heavy ion abundances in slow wind do not preferentially couple to any element.
In contrast, \As\ in Figure 5 of the \citetalias{ACE:SWICS:FSTransition} shows the fast solar wind dependence on FIP expected from \citet{Zurbuchen2016}, which is likely due to fractionation in the chromospheric and transition region.
\citet{ACE:SWICS:FSTransition} also report several unexpected results.
First, the saturation speed for elements heavier than \He\ is \vs[327 \pm 2], independent of mass.
From this, they infer that some process preferentially couples with \He, but not heavier elements.
Then, they show that \vs\ for helium is faster than heavy ion \vs\ by \kms[63].
From this, they infer that \He\ is impacted above the sonic critical point by an acceleration mechanism in a manner that heavy ions are not.
Because the peak of the solar wind distribution at \au[1] separates the \He\ and heavy \vs\ speeds, this observation may also be consistent with heavy elements being drawn out of the corona by collisional coupling with \Hy, but \He\ being accelerated in a different fashion.
Finally, they report a heavy ion fractionation in fast solar wind that depends on average heavy ion charge state (\Q) or (to a less extent) heavy ion mass (\M), but nor mass-per-charge ratio (\MpQ).
However, they do not provide an explanation for this.

In this Letter, we utilize the results of \citetalias{ACE:SWICS:FSTransition} to split heavy ion abundances into solar wind predominantly from magnetically closed regions (slow solar wind) and magnetically open regions (fast wind).
We then characterize how \ahe\ and heavy ion abundances of \He, \C, \N, \Ox, \Ne, \Mg, \Si, \Su\, and \Fe\ normalized to \Hy\ evolve with solar activity as quantified by the 13-month smoothed sunspot number \citep{SIDC} in the same manner as has been applied to \ahe\ \citep{Aellig:Ahe,Kasper:Ahe,Kasper:Ahe:Qstate,McIntosh:Ahe,Alterman2019,Alterman2021}.
To account for how the amplitude of \SSN\ changes across solar cycles, we have calculated a normalized \SSN\ (\NSSN) that scales \SSN\ in each activity cycle to its maximum value \citep{Zhao2013}.
By splitting each element into intervals that are faster and slower than its saturation speed, we are also able to extend the analysis of \ahe\ to faster speeds.
We report that \AbSW* and \ahe\ from slow solar wind vary with solar activity.
The variation of \ahe\ in fast solar wind with \SSN\ is also strong, but the correlation coefficients between fast wind \AbSW* and \SSN\ are insufficiently significant to draw inferences.
Comparing \ahe\ observed by SWICS and SWE, we cannot rule out that low p-values derived with SWICS observations in fast wind may be due to the limited time period of observations.
For the slow solar wind, the correlation between \SSN\ and \AbSW* for elements heavier than \He\ monotonically increases with increasing mass.
In particular, these slow wind correlation coefficients increase from 0.49 for \AbSW*[\C] to 0.81 for \AbSW*[\Fe], which is as strong as \AbSW*[\He].
We interpret the increase of this correlation coefficient with increasing element mass as a signature of gravitational settling while the strong correlation coefficient between \AbSW*[\He] and \SSN\ is more likely related to how helium is dynamically relevant for solar wind acceleration.

\section{Observations \label{sec:obs}}
We use the same dataset as \citet{ACE:SWICS:FSTransition}. 
As such, we combine observations of the helium abundance $\ahe = \AbSW[\He]$ from the Wind \citep{Wind} Solar Wind Experiment \citep[SWE,][]{Wind:SWE} Faraday cups (FC) and heavy ion observations from the Advanced Composition Explorer \citep[ACE,][]{ACE} Solar Wind Ion Composition Spectrometer \citep[SWICS,][]{ACE:SWICS}, a charge-resolving ion mass spectrometer.
We also use \Hy\ and \He\ observations available from the ACE Science Center \citep{ACE:ASC} using the same formulation as \citet{ACE:SWICS:AUX}.
A detailed discussion of the instruments are available in the cited papers.
Both ACE and Wind observations are exclusively collected in the ecliptic plane.
We use the 13-month smoothed sunspot number (\SSN) to trace solar activity \citep{SIDC}.
To calculate the normalized sunspot number (\NSSN), we subtract the minimum \SSN\ during each cycle and then normalize this shifted \SSN\ in a given cycle to its maximum value in that cycle.
For solar cycle 25, we use a maximum \SSN\ of 160.6, which is the prediction from NOAA's Space Weather Prediction Testbed\footnote{\url{https://testbed.spaceweather.gov/products/solar-cycle-progression-updated-prediction-experimental}}.

\section{Analysis \label{sec:analysis}}

\cref{fig:t-swe} plots the SWE helium abundance as a function of time.
The abundance has been split into fast and slow wind intervals, where the transition is defined by the speed \vs[420] at which the gradient of \ahe\ as a function of \vsw\ changes in \citetalias{ACE:SWICS:FSTransition}.
The interval over which SWICS observations are available is indicated at the bottom of the panel.
Over the full time period plotted, the correlation coefficients between the SWE abundances and \SSN\ are 0.94 (slow wind) and 0.66 (fast wind), both with p-values $< 0.05$.
Using \NSSN, we get 0.95 (slow wind) and 0.72 (fast wind) with p-values $< 0.05$.
When only considering the time period for which SWICS observations are available, the correlation coefficients change to 0.95 (slow wind) and 0.75 (fast wind), again with p-values $< 0.05$.
Using \NSSN, we get 0.95 (slow wind) and 0.76 (fast wind) with p-values $< 0.05$.
These are summarized in \cref{tbl:corr}.

In \cref{fig:t-swe}, the horizontal dotted line indicates \ahe[51 \pm 3] with respect to the photospheric helium abundance, the weighted mean of the fast wind observations in the 250-day intervals.
Here, the weights are calculated as the standard deviation of \ahe\ in each 250-day interval and the gray band indicates the weighted uncertainty.
The fast wind observations, which are predominantly from continuously open source regions, oscillate around half of the photospheric \ahe.
Slow wind \ahe, which is predominantly from intermittently open source regions, oscillates around a much larger range of values and reaches a maximum of this 51\% level during solar maxima.

\plotSWICSOverTime*
\cref{fig:t-swics} plots the SWICS abundances as a function of time in (a) fast and (b) slow wind.
In \cref{fig:t-swe}, we calculate \ahe\ as the simple average over the interval.
In \cref{fig:t-swics}, we calculate the weighted mean and standard error of the weighted mean because SWICS observations are reported with a 2hr cadence instead of 92s by SWE.
\ahe\ derived from SWE observations is normalized to its photospheric value and plotted to show consistency with SWICS \AbSW[\He].
The right axes indicate the 13-month smoothed SSN.
The vertical dotted lines indicate solar minimum 24.
The vertical dash-dotted line indicate the helium shutoff preceding solar minimum 24.
In general, there is a rough stratification in the abundances such that low FIP abundances are higher than high FIP abundances across the solar cycle.
With the exception of slow wind \Su\ and fast wind \Si, the observed minimum of fast wind \AbSW* precedes solar minimum 24 by 122 days and follows helium shutoff by 62 days.

\TblCorr
To characterize the evolution of these abundances with solar activity, we have calculated the correlation coefficient between each abundance and \NSSN.
For the SWE abundance, this is limited to the same time period over which SWICS abundances are available.
We have also calculated the correlation coefficient between all SWICS abundances and the SWE abundance.
These correlation coefficients are calculated for slow wind (\vsw*[\vs][<]) and fast wind (\vsw*[\vs][>]), where \vs\ is defined as in \citetalias{ACE:SWICS:FSTransition} for each species.
As the SWE and SWICS data have a different observation cadence, the boundaries of each data set's 250-day intervals do not necessarily overlap.
As such, the SWE data has been interpolated to the SWICS observation time for this calculation.
\cref{tbl:corr} summarizes these correlation coefficients for both slow and fast wind.

The \emph{SWE} columns indicates the correlation coefficients between \AbSW* and \ahe.
The \emph{SSN} columns indicate the correlation coefficients between a given abundance and \SSN.
Similarly, the \emph{NSSN} column indicates the correlation coefficients between a given abundance and \NSSN.
Only significant correlation coefficients (p-value $< 0.05$) are shown.
The correlation coefficient between fast and slow wind SWE abundances and \SSN\ is stronger than the correlation between SWICS abundances and \SSN.
In general, the correlation coefficients derived for slow wind observations increase with increasing mass and, excluding \C, all are strong ($> 0.6$ for \SSN\ and $> 0.7$ for \NSSN).
In the fast wind, only 4 (\SSN) or 5 (\NSSN) of 9 correlation coefficient using SWICS data are significant and none are strong ($< 0.6$); the exception is the correlation coefficient between \AbSW*[\Fe] and \NSSN, which exceeds 0.6 and is significant.
With the exception of slow wind \Fe, SWICS \AbSW*[\He] is larger than all other \AbSW* correlation coefficients with \SSN.
Slow wind \Fe's correlation coefficient with \SSN\ is larger to slow wind \He; its correlation coefficient with \NSSN\ in slow wind is equal to \He's.
In fast wind, \Fe\ exhibits the strongest correlation coefficient with \SSN\ and \NSSN.
In the slow and fast wind, \Fe\ exhibits the strongest correlation with \ahe\ observed by SWE.

\plotChemistryFinal
\cref{fig:chemistry-delay} plots the slow wind correlation coefficients with \NSSN\ (solid line) and \SSN\ (dash-dotted line) as a function of element mass (\M).
Markers are connected to aid the eye.
The top axis identifies each species in the color corresponding to the marker and a vertical dotted line in the same color connects this label to the marker.
The SWE abundance is labeled to differentiate it from the SWICS \AbSW[\He].
As in \cref{tbl:corr}, correlation coefficients using \NSSN\ are stronger than those using \SSN.
For elements heavier than \He, we also observe a monotonic increase in the correlation's strength with increasing mass.
An ordering by charge state (\Q), \MpQ, and FIP is less apparent.

\section{Discussion \label{sec:disc}}

Elemental abundances are set in the chromosphere and transition region.
Because such quantities are fixed by the time the coronal plasma enters the solar wind, they serve as tracers of solar wind source regions and the physical processes active at them.
As such, the evolution of these abundances with solar activity is driven by the evolution of their solar sources.
This work applies methods long used to study the evolution of the solar wind helium abundances with solar activity \citep{Aellig:Ahe,Kasper:Ahe,Kasper:Ahe:Qstate,Alterman2019,Alterman2021,Yogesh:Ahe} to heavy ion composition observations.
To achieve sufficient statistical reliability in the observations, we do not split the observations into 10 quantiles of \vsw.
Rather, we have split the solar wind into ``slow'' and ``fast'' regimes, where the different regimes are defined by speeds slower and faster than each species' \vs.
This ``saturations speed'' (\vs) is the speed at which the gradient of a given abundance as a function of \vsw\ changes.
The \citetalias{ACE:SWICS:FSTransition} calculates \vs\ for heavy ion abundances observed by ACE/SWICS.
When comparing \vs\ for heavy abundances and \ahe\ we will note \vs_\mathrm{He}_ and \vs_\He_, respectively, to differentiate the two.
The \citetalias{Wind:SWE:ahe:xhel} applies the same method to Wind/FC observations of \ahe.
Based on the evolution of \vs\ and the abundance at this speed (the saturation abundance \As) with normalized cross helicity, the \citetalias{Wind:SWE:ahe:xhel} argues that ``fast'' wind (\vsw*[\vs][>]) is from primarily polar solar source regions with magnetic topologies that are continuously open to the heliosphere and ``slow'' wind (\vsw*[\vs][<]) is from primarily equatorial solar source regions with magnetic topologies that are only intermittently open to the heliosphere.
The \citetalias{ACE:SWICS:FSTransition} shows that 
\begin{inparaenum}[(1)]
\item the gradient of \AbSW* and \ahe\ for speeds \vsw*[\vs][<] are indistinguishable, 
\item the heavy element saturation speeds is \vs[327 \pm 2]_\mathrm{Heavy}_ across all elements heavier than \He, 
\item $\vs_\mathrm{Heavy}_ > \vs_\He_$ by \kms[63 \pm 4.5], and 
\item there is a mass- or charge-state- dependent fractionation of \AbSW* for speeds \vsw*[\vs][>].
\end{inparaenum}
From the first observation, \citeauthor{ACE:SWICS:FSTransition} infer that heavy element and \He\ abundances are driven by the same process over speeds \vsw*[\vs][<].
From the second and third observations, they infer that the process impacting \AbSW* and \ahe\ for speeds \vsw*[\vs][<] happens below heights at which \vsw\ is established, therefore likely in the chromosphere and/or transition region.
They argue that these are both consistent with gravitational settling.
From the fourth observation, they identify a mass- or charge-state- dependent fractionation process that becomes increasingly significant with increasing speed, but provide no justification for this empirical observation.

\subsection{SWE Abundance \label{sec:disc:swe}}

\cref{fig:t-swe} plots the solar wind helium abundance in two speed intervals, faster and slower than \vs.
The correlation coefficient between \ahe\ and \SSN\ in both speed intervals is high ($> 0.6$ or better).
\citet{Alterman2019} analyzed \ahe's evolution with \SSN\ across 10 speed quantiles.
Data in each quantile is aggregated into 250-day intervals.
The slow wind interval defined in this work covers their 7 slowest speed quantiles.
Over these quantiles, the correlation coefficient between \ahe\ and \SSN\ exceeds 0.8.
The fast wind interval defined in this work covers \citepossessive{Alterman2019} three fastest speed intervals along with faster wind, extending the range of this analysis from a maximum of \vsw[600][\approx] to \vsw[800][\approx] \citep{Aellig:Ahe,Kasper:Ahe,Kasper:Ahe:Qstate,Alterman2019,Alterman2021}.
In these faster speed quantiles, the correlation coefficient between \ahe\ and \SSN\ markedly drops to below 0.7.
These correlation coefficients likely exceed those reported by \citet{Kasper:Ahe} across all speeds and those reported by \citet{Kasper:Ahe:Qstate} in faster quantiles because longer time intervals reduce the impact of fluctuations when calculating the correlation coefficients between two time series.
This is, for example, why the two solar cycles of activity reported by \citet{Alterman2019} enable them characterise the \vsw-dependent phase lag in \ahe's response to changes in \SSN.
As such, although the speed intervals reported in this paper are markedly larger than those reported by \citet{Alterman2019,Kasper:Ahe,Kasper:Ahe:Qstate} and speeds \vsw*[\vs][<] aggregate over the speed-dependent phase lag between \ahe\ and \SSN\ \citep{Alterman2019}, the larger speed intervals in this paper are less sensitive to fluctuations in the aggregated values.
This is likely why the correlation coefficient between \ahe\ and \SSN\ for speeds \vsw*[\vs][<] is closest to the strongest correlation coefficients reported for observations with speeds near the peak of the distribution of \vsw\ observed near \au[1] \citep{Kasper:Ahe,Kasper:Ahe:Qstate,Alterman2019}.
Regarding the speeds \vsw*[\vs][>], the probability density of the distribution of \vsw\ observed near \au[1] drops by nearly an order of magnitude from peak over the speed range 343 to \kms[\sim650].
Further dividing solar wind observations into smaller speed intervals increases the significance of random fluctuations in aggregated quantities.
By aggregating across a wide range of speeds, we reduce the impact of such statistical fluctuations, which is why we are able to determine that \ahe\ with speeds \vsw*[\vs][>] and therefore predominantly from source regions that are continuously open to the heliosphere also evolves with solar activity.
That the correlation coefficients derived with \NSSN\ exceed those derived with \SSN\ in both slow and fast wind indicates that the sunspot number is tracking the evolution of solar activity, not necessarily the physical process driving the evolution of \ahe\ with it.

The photospheric helium abundance is $8.25 \pm 0.2\%$ \citep{Asplund2021}.
For speeds \vsw*[\vs][>], solar wind \ahe\ oscillates around $51 \pm 0.3\%$ this value.
During solar maximum when solar wind sources like coronal holes with continuously open magnetic topologies are not confined to the Sun's polar regions, solar wind \ahe\ with speeds \vsw*[\vs][<] reaches a maximum during of 47.7\% (maximum 24) to 49.7\% (maximum 23) of its photospheric value.
Given that coronal mass ejections (CMEs) release solar plasma from deep within the solar atmosphere and are known to be enhanced in heavy element abundances and \ahe\ \citep[e.g.][]{Yogesh:CME,Rivera:CME,Song2022} and that slow wind \ahe\ is depleted from its fast wind values, this suggests that the mechanism limiting non-transient solar wind \ahe\ to $\sim50\%$ of its photospheric value occurs at heights near or below the chromosphere and transition region.

\subsection{SWICS Abundances \label{sec:disc:swics}}
Per the \citetalias{Wind:SWE:ahe:xhel} and the \citetalias{ACE:SWICS:FSTransition}, solar wind observations at \au[1] with \vsw*[\vs][>] are predominantly from source regions with magnetic field topologies that are continuously open to the heliosphere.
In contrast, \vsw*[\vs][<] identifies solar wind that is predominantly from intermittently open source regions.
We have divided \AbSW* observations into intervals with speeds above and below \vs, calculated \vs\ independently for each species \citepalias{ACE:SWICS:FSTransition}.
\cref{fig:t-swics} plots each \AbSW* as a function of time aggregated in 250-day intervals.
In this plot, we label \vsw*[\vs][>] as ``fast wind'' and \vsw*[\vs][\leq] ``slow wind''.
As expected, low FIP abundances are higher than high FIP abundances across the solar cycle.

Per \cref{tbl:corr}, the correlation coefficients between \SSN\ and these abundances is strong ($>0.6$) and significant (p-value $< 0.05$) for all species in solar wind from intermittently open source regions.
The \citetalias{ACE:SWICS:FSTransition} shows that the gradient of \AbSW* as a function of \vsw\ for \vsw*[\vs][<] are consistent across all species and infers that the underlying physical mechanism does not preferentially couple to any element.
With the exception of \He, \cref{fig:chemistry-delay} shows that the correlation coefficient between \SSN\ and \AbSW* with slow speeds from intermittently open source regions monotonically increases with increasing \M.

It is unsurprising that \He\ does not follow the trend in \cref{fig:chemistry-delay}.
The \citetalias{Wind:SWE:ahe:xhel} argues that \He\ is dynamically relevant for solar wind acceleration in both continuously and intermittently open source regions.
This is likely why the correlation coefficient between \ahe\ and \SSN\ is strong and significant for both the \vsw*[\vs][<] and \vsw*[\vs][>] speed ranges. 
In contrast, heavy element abundances are likely too small to be dynamically relevant to the solar plasma's overall acceleration into the solar wind.

The similarity in the correlation coefficients of \SSN\ with \AbSW* and \ahe\ for \vsw*[\vs][<] suggests that one or more processes drive the long term evolution of these abundances in similar ways.
This is reflected in the correlation coefficients between \ahe\ and \AbSW* in that they are, with the exception of \C, strong and significant in slow wind with \vsw*[\vs][<] and either weak ($< 0.6$) or not significant (high p-value $> 0.05$) for \vsw*[\vs][>].
That the correlation coefficients derived with \NSSN\ exceed \SSN\ further suggests the sunspot number is a ``clock'' for timing solar activity, but doesn't trace the underlying process driving the evolution of \AbSW*.

Gravitational settling is one process that governs abundances in closed loops that is coupled to elements based on their mass, not charge state, FIP, $Q/M$, etc.\ \citep{Vauclair1991,Borrini1981,Hirshberg1973a,Weberg2015a,Alterman2019}.
Such loops are longer lived during solar minima, leading to more depleted abundances and such a mechanism does not affect solar wind where such closed loops are absent.
As such, we infer that the difference between the evolution with solar activity of solar wind abundances from continuously and intermittently open source regions may be driven by gravitational settling in such regions.

That the correlation coefficient between slow wind \ahe\ and \SSN\ is higher than the heavier elements may signify that helium's evolution with solar activity is driven by additional processes \citepalias[e.g. the][and references there in]{Wind:SWE:ahe:xhel} that do not couple to heavier elements.
Figure 7 in the \citetalias{ACE:SWICS:FSTransition} shows that $\vs_\He_ = \vs_\mathrm{Heavy}_ + \kms[63]$ and these two speeds are separated by the peak of the solar wind speed distribution, further suggesting that some other process is accelerating \He\ to faster speeds than heavy elements.
However, we cannot rule out that the 2-hr duration of SWICS measurements and means that we cannot separate solar wind by speed when \vsw*[\vs][\sim] for SWICS observations with the same precision as we can for SWE observations and that only one solar cycle of SWICS observations means that statistical fluctuations have a more significant impact on our SWICS observations than our SWE observations for \vsw*[\vs][>]

\subsection{Composition Shutoff \label{sec:disc:shutoff}}
The time between the observed minimum in fast wind \AbSW* and the helium shutoff preceding solar minimum 24 \citep{Alterman2021} is half as long as the time between the minimum in fast wind \AbSW* and solar minimum 24.
However, both of these intervals are within the 250-day averaging window and the 229 day uncertainty in the helium shutoff.
As such, any inference from this observation requires higher time resolution analysis.

\subsection{Drivers of the Helium and Heavy Element Abundances \label{sec:disc:sources}}

\citet{McIntosh:Ahe} study how the relationship between \ahe, the iron-to-oxygen ratio (\AbSW[\Fe][\Ox]), average iron charge state (\Q*[\Fe]), and characteristic scale size in the transition region evolve with solar activity in slow (\vsw[400][<]) and fast (\vsw[500][>]) solar wind.
They argue that the drop in \ahe\ during solar minima is consistent with a decrease in the energy available to accelerate coronal helium into the solar wind during this phase of solar activity.
They suggest that this is consistent with a decrease in the transition regions scale size during solar minima, which also implies less energy available to heat the solar wind at these heights, resulting in a reduced number density \citep{McComas2008} and \Q*[\Fe].
This depletion in available energy is consistent with observations of solar wind \Qratio{\Ox}{7}{6} and \Qratio{\C}{6}{5} charge state ratios that evolve with \SSN\ and drop during solar minima \citep{Kasper:Ahe:Qstate}.
Given that coronal electron temperatures and densities likely do not change with solar activity \citep{Landi2014} and that the efficiency of solar wind acceleration likely varies with source regions, the evolution of these parameters with solar activity likely reflects the evolution of the frequency at which solar wind from different source regions is sampled at \au[1].

In the case of fast wind, the solar wind originates from continuously open source regions with radial magnetic fields.
In such solar wind, \ahe\ and heavy ion abundances do not vary significantly because the solar wind primarily originates from a single class of source region.

In the case of slow wind \ahe, \citet{Kasper:Ahe:Qstate} attribute its variability to the occurrence of two different sources.
One is active regions (ARs), the edge of which recent studies suggest may be consistent with sources of the Alfvénic slow wind \citep{Baker2023,Yardley2024,Ervin2024}.
The other is the streamer belt.
\citet{Alterman2019} argue that the \vsw-dependent phase lag in \ahe's response to changes in \SSN\ for speeds \vsw[574][\leq] is consistent with a \He\ filtration mechanism.
When contextualized with the results of the \citetalias{Wind:SWE:ahe:xhel}, such a filtration mechanism may be related to the energy available to accelerate coronal helium into the solar wind, which varies with the source region \citep{Hansteen1997,Endeve2005}.
Such an inference is consistent with \citet{Yogesh:Ahe}, who argue that this filtration mechanism is related to the topology of the Sun's magnetic field.
If the heat flux mediated coupling between the corona and transition region is significant for determining the energy available to accelerate the solar wind or, equivalently, the acceleration efficiency \citep{Lie-Svendsen2001,Lie-Svendsen2002}, such an inference is also consistent with differences between fast wind being, ``generated in the low corona, where \C\ and \Ox\ freeze-in, and then the two winds experience a similar evolution in the extended corona, where \Fe\ freezes-in \citep{ACE:SWICS:AUX}.''
If this is the case, the highly variable \AbSW\ in solar wind with \vsw*[\vs][<] also reflects the energy available to accelerate heavy ions from the solar plasma into the solar wind.
The \vsw-dependent filtration mechanism may be related to helium's dynamic role in solar wind acceleration \citep{Hansteen1997,Endeve2005,Wind:SWE:ahe:xhel}, for which heavy element abundances are too small.

\section{Conclusion \label{sec:conclusion}}

We have observed the evolution of the solar wind helium abundance and heavy element abundances with solar activity.
The helium abundance observations cover approximately 22 years of Wind/SWE operations, while the abundances of heavier elements are limited to approximately 12 years during ACE/SWICS operations prior to the detector degradation.
Using the solar wind helium abundance, we have made the following observations.
\begin{compactenum}
\item \ahe\ observed in slow and fast wind correlates strongly with the Sun's activity cycle as observed in \SSN.
\item The correlation coefficients derived with the normalized \SSN\ (\NSSN) are stronger than those derived with \SSN, indicating that the sunspot number is a ``clock'' timing the evolution of solar activity, not the underlying physical process that modulates \ahe.
\item Solar wind helium in fast wind, which is predominantly from continuously open magnetic sources,  oscillates around $51\%$ of its photospheric value.
\item In slow wind from intermittently open source regions, \ahe\ is highly variable below $51\%$ and only reaches this maximum value during solar maxima when continuously and intermittently open source regions are not confined to the Sun's polar regions.
\end{compactenum}

Using heavy ion abundances from ACE/SWICS, we have made the following observations.
\begin{compactenum}
\item In slow wind, the abundance of all heavy elements except \C\ is strongly correlated with \SSN.
\item In fast wind, \AbSW* abundances do not evolve significantly with solar activity.
However, the more periods of oscillation observed, the less significant the impact any random fluctuations would have on the correlation coefficient.
Because
\begin{inparaenum}[(a)]
\item we have more than two solar activity cycles of \ahe\ observations and just more than one solar activity cycle of \AbSW* observations and 
\item fast wind oscillations are smaller in amplitude than slow wind oscillations, 
\end{inparaenum}
we cannot conclusively state that this is because \AbSW* in fast wind does not evolve with solar activity or if the result is due to the limited duration of our observations.
\item The correlation coefficients derived between \AbSW* and \NSSN\ are stronger than those derived with \SSN, further indicating that the sunpot number is tracking solar activity, but not the underlying process that drives the evolution of the abundances.
\item Unsurprisingly, the strong correlation between \ahe\ and \SSN\ along with \AbSW*\ and \SSN\ is reflected in a strong correlation between \ahe\ and \AbSW*.
That these correlation coefficients are weaker is unsurprising because we use the 13-month smoothed \SSN\ and neither the 250-day average \ahe\ nor \AbSW* are smoothed.
\item The correlation coefficients between \AbSW*\ and \SSN\ monotonically increase with increasing element mass.
We infer that this is a signature of gravitational settling in slow wind, which is consistent with the gradient of \AbSW* over speeds \vsw*[\vs][<] being indistinguishable across the heavier elements \citepalias{ACE:SWICS:FSTransition}.
\item The helium shutoff is a precipitous depletion and recovery in \ahe\ that precedes \SSN\ minima by approximately 229 days \citep{Alterman2021}.
The largest source of uncertainty on this observation is the 250-day averaging window.
The minimum \AbSW* we observed precedes \SSN\ minima by 122 days, but follows the helium shutoff by only 62 days.
In other words, the time between the minimum in \AbSW* and the helium shutoff is half as long as the time between the minimum \AbSW* and \SSN\ minima.
However, the difference between these two times is smaller than the 229 day uncertainty on helium shutoff and the 250-day averaging window.
As such, higher time resolution analysis is necessary to determine if the depletion of \AbSW* during solar minima coincides with helium shutoff.
\end{compactenum}

We attribute the differences between the evolution of \ahe\ and \AbSW* in slow wind with solar activity to helium's dynamic involvement in solar wind acceleration \citepalias[and references therein]{Wind:SWE:ahe:xhel}.
These differences are also reflected in the difference between \vs\ for \He\ and heavier elements derived in the \citetalias{ACE:SWICS:FSTransition}.
The differences in fast wind are either also related to how helium is dynamically relevant for solar wind acceleration or simply driven by the lower SWICS data volume with respect to SWE.
More broadly, we infer that the variation in both \ahe\ and \AbSW* with solar activity is driven by the energy in the chromosphere and transition region or low corona, which is also related to the topology of the magnetic field in the solar wind's source regions.
This suggests that the evolution of slow wind \ahe\ and \AbSW* with solar activity observed at \au[1] may be due to changes in the frequency of observing solar wind from different source regions.

In closing, we note that both ACE's and Wind's orbit are in the ecliptic plane.
Given that the occurrence rate and heliographic latitude of different solar source regions changes with solar activity, long duration out of ecliptic observations are necessary to fully understand the relationship between source region and in situ abundances.
Furthermore, recent results show that Alfvén wave energy deposition and thermal pressure gradients accelerate the solar wind in transit through interplanetary space \citep{Rivera2024,Rivera2025}.
This means such out of ecliptic observations need to be either collected over a long duration and at a single radial distance or the relevant in situ acceleration mechanisms must be sufficiently well understood that radial gradients in \vsw\ can be properly accounted for, e.g., with observations from Ulysses/SWICS \citep{Ulysses:SWICS,vonSteiger2000}, so that we can properly trace in situ observations back to their source regions.
Such observations are a critical component of future missions \citep{Rivera2025a}.

\begin{acknowledgements}
BLA thanks F. Carcaboso for helpful discussions about the helium abundance variation with time.
BLA acknowledges funding from NASA Grants 80NSSC22K0645 (LWS/TM) and 80NSSC22K1011 (LWS).
JMR and STL acknowledge NASA contract
80NSSC23K0542 (ACE/SWICS).
Sunspot data from the World Data Center SILSO, Royal Observatory of Belgium, Brussels.
The authors thank the referee for their useful questions and helpful suggestions.
\end{acknowledgements}

\bibliography{Zotero.bib}{}

\begin{thebibliography}{90}
\expandafter\ifx\csname natexlab\endcsname\relax\def\natexlab#1{#1}\fi

\bibitem[{Abbo {et~al.}(2016)Abbo, Ofman, Antiochos, Hansteen, Harra, Ko,
  Lapenta, Li, Riley, Strachan, von Steiger, \& Wang}]{Abbo2016}
Abbo, L., Ofman, L., Antiochos, S.~K., {et~al.} 2016, Space Science Reviews,
  201, 55, tex.ids= Abbo2016a

\bibitem[{Acuña {et~al.}(1995)Acuña, Ogilvie, Baker, Curtis, Fairfield, \&
  Mish}]{Wind}
Acuña, M.~H., Ogilvie, K.~W., Baker, D.~N., {et~al.} 1995, Space Science
  Reviews, 71, 5

\bibitem[{Aellig {et~al.}(2001)Aellig, Lazarus, \& Steinberg}]{Aellig:Ahe}
Aellig, M.~R., Lazarus, A.~J., \& Steinberg, J.~T. 2001, Geophysical Research
  Letters, 28, 2767

\bibitem[{Alterman \& D'Amicis(2025)}]{Wind:SWE:ahe:xhel}
Alterman, B.~L. \& D'Amicis, R. 2025, The Astrophysical Journal

\bibitem[{Alterman \& Kasper(2019)}]{Alterman2019}
Alterman, B.~L. \& Kasper, J.~C. 2019, The Astrophysical Journal, 879, L6,
  publisher: IOP Publishing

\bibitem[{Alterman {et~al.}(2021)Alterman, Kasper, Leamon, \&
  McIntosh}]{Alterman2021}
Alterman, B.~L., Kasper, J.~C., Leamon, R.~J., \& McIntosh, S.~W. 2021, Solar
  Physics, 296, 67, arXiv: 2006.04669 Publisher: The Author(s), under exclusive
  licence to Springer Nature B.V. ISBN: 1120702101801

\bibitem[{Alterman {et~al.}(2018)Alterman, Kasper, Stevens, \&
  Koval}]{Alterman2018}
Alterman, B.~L., Kasper, J.~C., Stevens, M., \& Koval, A. 2018, The
  Astrophysical Journal, 864, 112, publisher: IOP Publishing

\bibitem[{Alterman {et~al.}(2025)Alterman, Rivera, Lepri, \&
  Raines}]{ACE:SWICS:FSTransition}
Alterman, B.~L., Rivera, Y.~J., Lepri, S.~T., \& Raines, J.~M. 2025, Astronomy
  \& Astrophysics

\bibitem[{Antiochos {et~al.}(2011)Antiochos, Mikic, Titov, Lionello, \&
  Linker}]{Antiochos2011}
Antiochos, S.~K., Mikic, Z., Titov, V.~S., Lionello, R., \& Linker, J.~A. 2011,
  The Astrophysical Journal, 112, arXiv: 1102.3704

\bibitem[{Antonucci {et~al.}(2005)Antonucci, Abbo, \& Dodero}]{Antonucci2005}
Antonucci, E., Abbo, L., \& Dodero, M.~A. 2005, Astronomy \& Astrophysics, 435,
  699

\bibitem[{Asplund {et~al.}(2021)Asplund, Amarsi, \& Grevesse}]{Asplund2021}
Asplund, M., Amarsi, A.~M., \& Grevesse, N. 2021, Astronomy \& Astrophysics,
  653, A141

\bibitem[{Baker {et~al.}(2023)Baker, Démoulin, Yardley, Mihailescu, Van
  Driel-Gesztelyi, D’Amicis, Long, To, Owen, Horbury, Brooks, Perrone,
  French, James, Janvier, Matthews, Stangalini, Valori, Smith, Cuadrado, Peter,
  Schuehle, Harra, Barczynski, Berghmans, Zhukov, Rodriguez, \&
  Verbeeck}]{Baker2023}
Baker, D., Démoulin, P., Yardley, S.~L., {et~al.} 2023, The Astrophysical
  Journal, 950, 65

\bibitem[{Berger {et~al.}(2011)Berger, Wimmer-Schweingruber, \&
  Gloeckler}]{Berger2011}
Berger, L., Wimmer-Schweingruber, R.~F., \& Gloeckler, G. 2011, Physical Review
  Letters, 106, 151103

\bibitem[{Borrini {et~al.}(1981)Borrini, Gosling, Bame, Feldman, Wilcox,
  Gosling, Bame, \& Feldman}]{Borrini1981}
Borrini, G., Gosling, J.~T., Bame, S.~J., {et~al.} 1981, Journal of Geophysical
  Research: Space Physics, 86, 4565

\bibitem[{Brooks {et~al.}(2015)Brooks, Ugarte-Urra, \& Warren}]{Brooks2015}
Brooks, D.~H., Ugarte-Urra, I., \& Warren, H.~P. 2015, Nature Communications,
  6, publisher: Nature Publishing Group

\bibitem[{Bruno \& Carbone(2013)}]{LR:turbulence}
Bruno, R. \& Carbone, V. 2013, Living Reviews in Solar Physics, 10, 1

\bibitem[{Crooker {et~al.}(2012)Crooker, Antiochos, Zhao, \&
  Neugebauer}]{Crooker2012}
Crooker, N.~U., Antiochos, S.~K., Zhao, X., \& Neugebauer, M. 2012, Journal of
  Geophysical Research: Space Physics, 117, n/a, iSBN: 0148-0227

\bibitem[{D'Amicis \& Bruno(2015)}]{DAmicis2015}
D'Amicis, R. \& Bruno, R. 2015, Astrophysical Journal, 805, 1, publisher: IOP
  Publishing ISBN: 1538-4357

\bibitem[{D'Amicis {et~al.}(2016)D'Amicis, Bruno, \& Matteini}]{Damicis2016}
D'Amicis, R., Bruno, R., \& Matteini, L. 2016, AIP Conference Proceedings,
  1720, iSBN: 9780735413672

\bibitem[{D'Amicis {et~al.}(2018)D'Amicis, Matteini, \& Bruno}]{DAmicis2018}
D'Amicis, R., Matteini, L., \& Bruno, R. 2018, Monthly Notices of the Royal
  Astronomical Society, 14, 1, arXiv: 1812.01899 tex.ids= Damicis2019

\bibitem[{D’Amicis {et~al.}(2021{\natexlab{a}})D’Amicis, Alielden, Perrone,
  Bruno, Telloni, Raines, Lepri, \& Zhao}]{DAmicis2021}
D’Amicis, R., Alielden, K., Perrone, D., {et~al.} 2021{\natexlab{a}},
  Astronomy \& Astrophysics, 654, A111

\bibitem[{D’Amicis {et~al.}(2011)D’Amicis, Bruno, \&
  Bavassano}]{DAmicis2011a}
D’Amicis, R., Bruno, R., \& Bavassano, B. 2011, Journal of Atmospheric and
  Solar-Terrestrial Physics, 73, 653, publisher: Elsevier

\bibitem[{D’Amicis {et~al.}(2021{\natexlab{b}})D’Amicis, Perrone, Bruno, \&
  Velli}]{DAmicis2021a}
D’Amicis, R., Perrone, D., Bruno, R., \& Velli, M. 2021{\natexlab{b}},
  Journal of Geophysical Research: Space Physics, 126

\bibitem[{Elsasser(1950)}]{ElsasserVariables}
Elsasser, W.~M. 1950, Physical Review, 79, 183

\bibitem[{Endeve {et~al.}(2005)Endeve, Lie‐Svendsen, Hansteen, \&
  Leer}]{Endeve2005}
Endeve, E., Lie‐Svendsen, O., Hansteen, V.~H., \& Leer, E. 2005, The
  Astrophysical Journal, 624, 402

\bibitem[{Ervin {et~al.}(2024{\natexlab{a}})Ervin, Bale, Badman, Rivera, Romeo,
  Huang, Riley, Bowen, Lepri, \& Dewey}]{Ervin2024a}
Ervin, T., Bale, S.~D., Badman, S.~T., {et~al.} 2024{\natexlab{a}}, The
  Astrophysical Journal, 969, 83

\bibitem[{Ervin {et~al.}(2024{\natexlab{b}})Ervin, Jaffarove, Badman, Huang,
  Rivera, \& Bale}]{Ervin2024}
Ervin, T., Jaffarove, K., Badman, S.~T., {et~al.} 2024{\natexlab{b}}, The
  Astrophysical Journal, 975, 156

\bibitem[{Feldman {et~al.}(1978)Feldman, Asbridge, Bame, \&
  Gosling}]{Feldman1978}
Feldman, W.~C., Asbridge, J.~R., Bame, S.~J., \& Gosling, J.~T. 1978, Journal
  of Geophysical Research, 83, 2177, tex.ids= Feldman1978a

\bibitem[{Fisk {et~al.}(1999)Fisk, Zurbuchen, \& Schwadron}]{Fisk1999}
Fisk, L.~A., Zurbuchen, T.~H., \& Schwadron, N.~A. 1999, The Astrophysical
  Journal, 521, 868

\bibitem[{Fu {et~al.}(2015)Fu, Li, Li, Huang, Mou, Jiao, \& Xia}]{Fu2015}
Fu, H., Li, B., Li, X., {et~al.} 2015, Solar Physics, 290, 1399

\bibitem[{Fu {et~al.}(2017)Fu, Madjarska, Xia, Li, Huang, \& Wangguan}]{Fu2017}
Fu, H., Madjarska, M.~S., Xia, L., {et~al.} 2017, The Astrophysical Journal,
  836, 169

\bibitem[{Garrard {et~al.}(1998)Garrard, Davis, Hammond, \& Sears}]{ACE:ASC}
Garrard, T., Davis, A.~J., Hammond, J., \& Sears, S. 1998, Space Science
  Reviews, 86, 649, publisher: Kluwer Academic Publishers

\bibitem[{Geiss(1982)}]{Geiss1982a}
Geiss, J. 1982, Space Science Reviews, 33, 201, iSBN: 9783540773405

\bibitem[{Geiss {et~al.}(1995{\natexlab{a}})Geiss, Gloeckler, \& von
  Steiger}]{Geiss1995b}
Geiss, J., Gloeckler, G., \& von Steiger, R. 1995{\natexlab{a}}, Space Science
  Reviews, 72, 49

\bibitem[{Geiss {et~al.}(1995{\natexlab{b}})Geiss, Gloeckler, Von~Steiger,
  Balsiger, Fisk, Galvin, Ipavich, Livi, McKenzie, Ogilvie, Et, \&
  Wilken}]{Geiss1995}
Geiss, J., Gloeckler, G., Von~Steiger, R., {et~al.} 1995{\natexlab{b}},
  Science, 268, 1033, publisher: Physikalisches Institut, University of Bern,
  Switzerland.

\bibitem[{Gloeckler {et~al.}(1998)Gloeckler, Cain, Ipavich, Tums, Bedini, Fisk,
  Zurbuchen, Bochsler, Fischer, Wimmer-Schweingruber, Geiss, Kallenbach, \&
  Kallenback}]{ACE:SWICS}
Gloeckler, G., Cain, J., Ipavich, F.~M., {et~al.} 1998, Space Sci. Rev., 86,
  497, publisher: Kluwer Academic Publishers

\bibitem[{Gloeckler {et~al.}(1992)Gloeckler, Geiss, Balsiger, Bedini, Cain,
  Fischer, Fisk, Galvin, Gliem, Hamilton, Hollweg, Ipavich, Joos, Livi,
  Lundgren, Mall, McKenzie, Ogilvie, Ottens, Rieck, Tums, von Steiger, Weiss,
  \& Wilken}]{Ulysses:SWICS}
Gloeckler, G., Geiss, J., Balsiger, H., {et~al.} 1992, Astronomy and
  Astrophysics Supplement Series, 92, 267

\bibitem[{Grappin {et~al.}(1991)Grappin, Velli, \& Mangeney}]{Grappin1991}
Grappin, R., Velli, M., \& Mangeney, A. 1991, Annales Geophysicae, 9, 416,
  publisher: Gauthier-Villars

\bibitem[{Hansteen {et~al.}(1997)Hansteen, Leer, \& Holzer}]{Hansteen1997}
Hansteen, V.~H., Leer, E., \& Holzer, T.~E. 1997, The Astrophysical Journal,
  482, 498

\bibitem[{Hirshberg(1973)}]{Hirshberg1973a}
Hirshberg, J. 1973, Astrophysics and Space Science, 20, 473

\bibitem[{Kasper {et~al.}(2017)Kasper, Klein, Weber, Maksimovic, Zaslavsky,
  Bale, Maruca, Stevens, \& Case}]{Kasper2017}
Kasper, J.~C., Klein, K.~G., Weber, T., {et~al.} 2017, The Astrophysical
  Journal, 849, 126

\bibitem[{Kasper {et~al.}(2008)Kasper, Lazarus, \& Gary}]{Kasper2008}
Kasper, J.~C., Lazarus, A.~J., \& Gary, S.~P. 2008, Physical Review Letters,
  101, 261103

\bibitem[{Kasper {et~al.}(2006)Kasper, Lazarus, Steinberg, Ogilvie, \&
  Szabo}]{Wind:SWE:bimax}
Kasper, J.~C., Lazarus, A.~J., Steinberg, J.~T., Ogilvie, K.~W., \& Szabo, A.
  2006, Journal of Geophysical Research, 111, A03105

\bibitem[{Kasper {et~al.}(2007)Kasper, Stevens, Lazarus, Steinberg, \&
  Ogilvie}]{Kasper:Ahe}
Kasper, J.~C., Stevens, M., Lazarus, A.~J., Steinberg, J.~T., \& Ogilvie, K.~W.
  2007, The Astrophysical Journal, 660, 901

\bibitem[{Kasper {et~al.}(2012)Kasper, Stevens, Korreck, Maruca, Kiefer,
  Schwadron, \& Lepri}]{Kasper:Ahe:Qstate}
Kasper, J.~C., Stevens, M.~L., Korreck, K.~E., {et~al.} 2012, The Astrophysical
  Journal, 745, 162

\bibitem[{Klein {et~al.}(2021)Klein, Verniero, Alterman, Bale, Case, Kasper,
  Korreck, Larson, Lichko, Livi, McManus, Martinović, Rahmati, Stevens, \&
  Whittlesey}]{Klein2021}
Klein, K.~G., Verniero, J.~L., Alterman, B.~L., {et~al.} 2021, The
  Astrophysical Journal, 909, 7, arXiv: 2101.10937

\bibitem[{Laming(2004)}]{Laming2004}
Laming, J.~M. 2004, The Astrophysical Journal, 614, 1063

\bibitem[{Laming(2009)}]{Laming2009}
Laming, J.~M. 2009, The Astrophysical Journal, 695, 954

\bibitem[{Laming(2015)}]{LR:FIP}
Laming, J.~M. 2015, Living Reviews in Solar Physics, 12, arXiv: 1504.08325
  ISBN: 2367-3648

\bibitem[{Landi \& Testa(2014)}]{Landi2014}
Landi, E. \& Testa, P. 2014, The Astrophysical Journal, 787, 33

\bibitem[{Lepri {et~al.}(2013)Lepri, Landi, \& Zurbuchen}]{ACE:SWICS:AUX}
Lepri, S.~T., Landi, E., \& Zurbuchen, T.~H. 2013, The Astrophysical Journal,
  768, 94

\bibitem[{Lepri \& Rivera(2021)}]{Lepri2021}
Lepri, S.~T. \& Rivera, Y.~J. 2021, The Astrophysical Journal, 912, 51,
  publisher: IOP Publishing tex.ids= Lepri2021a, Lepri2021b

\bibitem[{Lie-Svendsen {et~al.}(2001)Lie-Svendsen, Leer, \&
  Hansteen}]{Lie-Svendsen2001}
Lie-Svendsen, O., Leer, E., \& Hansteen, V.~H. 2001, Journal of Geophysical
  Research: Space Physics, 106, 8217

\bibitem[{Lie‐Svendsen {et~al.}(2002)Lie‐Svendsen, Hansteen, Leer, \&
  Holzer}]{Lie-Svendsen2002}
Lie‐Svendsen, O., Hansteen, V.~H., Leer, E., \& Holzer, T.~E. 2002, The
  Astrophysical Journal, 566, 562

\bibitem[{Marsch {et~al.}(1981)Marsch, Mühlhäuser, Rosenbauer, Schwenn, \&
  Denskat}]{Marsch1981}
Marsch, E., Mühlhäuser, K.-H., Rosenbauer, H., Schwenn, R., \& Denskat, K.~U.
  1981, Journal of Geophysical Research, 86, 9199

\bibitem[{McComas {et~al.}(2008)McComas, Ebert, Elliott, Goldstein, Gosling,
  Schwadron, \& Skoug}]{McComas2008}
McComas, D.~J., Ebert, R.~W., Elliott, H.~A., {et~al.} 2008, Geophysical
  Research Letters, 35, L18103

\bibitem[{McIntosh {et~al.}(2011)McIntosh, Kiefer, Leamon, Kasper, \&
  Stevens}]{McIntosh:Ahe}
McIntosh, S.~W., Kiefer, K.~K., Leamon, R.~J., Kasper, J.~C., \& Stevens, M.
  2011, Astrophysical Journal Letters, 740, 1, arXiv: 1109.1408

\bibitem[{Ogilvie {et~al.}(1995)Ogilvie, Chornay, Fritzenreiter, Hunsaker,
  Keller, Lobell, Miller, Scudder, Sittler, Torbert, Bodet, Needell, Lazarus,
  Steinberg, Tappan, Mavretic, \& Gergin}]{Wind:SWE}
Ogilvie, K.~W., Chornay, D.~J., Fritzenreiter, R.~J., {et~al.} 1995, Space
  Science Reviews, 71, 55

\bibitem[{Phillips {et~al.}(1994)Phillips, Balogh, Bame, Goldstein, Gosling,
  Hoeksema, McComas, Neugebauer, Sheeley, \& Wang}]{Phillips1994}
Phillips, J.~L., Balogh, A., Bame, S.~J., {et~al.} 1994, Geophysical Research
  Letters, 21, 1105

\bibitem[{Rakowski \& Laming(2012)}]{Rakowski2012}
Rakowski, C.~E. \& Laming, J.~M. 2012, The Astrophysical Journal, 754, 65,
  arXiv: 1204.2776v1

\bibitem[{Rivera \& Badman(2025)}]{Rivera2025a}
Rivera, Y.~J. \& Badman, S.~T. 2025, An assessment of observational coverage
  and gaps for robust {Sun} to heliosphere integrated science, arXiv:2502.06036
  [astro-ph]

\bibitem[{Rivera {et~al.}(2024)Rivera, Badman, Stevens, Verniero, Stawarz, Shi,
  Raines, Paulson, Owen, Niembro, Louarn, Livi, Lepri, Kasper, Horbury,
  Halekas, Dewey, De~Marco, \& Bale}]{Rivera2024}
Rivera, Y.~J., Badman, S.~T., Stevens, M.~L., {et~al.} 2024, Science, 385, 962

\bibitem[{Rivera {et~al.}(2025)Rivera, Badman, Verniero, Varesano, Stevens,
  Stawarz, Reeves, Raines, Raymond, Owen, Livi, Lepri, Landi, Halekas, Ervin,
  Dewey, De~Marco, D’Amicis, Dakeyo, Bale, \& Alterman}]{Rivera2025}
Rivera, Y.~J., Badman, S.~T., Verniero, J.~L., {et~al.} 2025, The Astrophysical
  Journal, 980, 70

\bibitem[{Rivera {et~al.}(2022{\natexlab{a}})Rivera, Higginson, Lepri, Viall,
  Alterman, Landi, Spitzer, Raines, Cranmer, Laming, Mason, Wallace, Raymond,
  Lynch, Gilly, Chen, \& Dewey}]{Rivera2022a}
Rivera, Y.~J., Higginson, A., Lepri, S.~T., {et~al.} 2022{\natexlab{a}},
  Frontiers in Astronomy and Space Sciences, 9, 1056347

\bibitem[{Rivera {et~al.}(2022{\natexlab{b}})Rivera, Raymond, Landi, Lepri,
  Reeves, Stevens, \& Alterman}]{Rivera:CME}
Rivera, Y.~J., Raymond, J.~C., Landi, E., {et~al.} 2022{\natexlab{b}}, The
  Astrophysical Journal, 936, 83

\bibitem[{Schwadron {et~al.}(1999)Schwadron, Fisk, \&
  Zurbuchen}]{Schwadron1999}
Schwadron, N.~A., Fisk, L.~A., \& Zurbuchen, T.~H. 1999, The Astrophysical
  Journal, 521, 859

\bibitem[{{SILSO World Data Center}(2023)}]{SIDC}
{SILSO World Data Center}. 2023, place: Royal Observatory of Belgium, avenue
  Circulaire 3, 1180 Brussels, Belgium

\bibitem[{Song {et~al.}(2022)Song, Cheng, Li, Zhang, \& Chen}]{Song2022}
Song, H., Cheng, X., Li, L., Zhang, J., \& Chen, Y. 2022, The Astrophysical
  Journal, 925, 137

\bibitem[{Stakhiv {et~al.}(2016)Stakhiv, Lepri, Landi, Tracy, \&
  Zurbuchen}]{Stakhiv2016}
Stakhiv, M.~O., Lepri, S.~T., Landi, E., Tracy, P.~J., \& Zurbuchen, T.~H.
  2016, The Astrophysical Journal, 829, 117, publisher: IOP Publishing ISBN:
  0769518745

\bibitem[{Stone {et~al.}(1998)Stone, Frandsen, Mewaldt, Christian, Margolies,
  Ormes, \& Snow}]{ACE}
Stone, E.~C., Frandsen, A.~M., Mewaldt, R.~A., {et~al.} 1998, Space Science
  Reviews, 86, 1, publisher: Kluwer Academic Publishers ISBN:
  10.1023/A:1005082526237

\bibitem[{Subramanian {et~al.}(2010)Subramanian, Madjarska, \&
  Doyle}]{Subramanian2010}
Subramanian, S., Madjarska, M.~S., \& Doyle, J.~G. 2010, Astronomy and
  Astrophysics, 516, A50

\bibitem[{Tracy {et~al.}(2016)Tracy, Kasper, Raines, Shearer, Gilbert, \&
  Zurbuchen}]{Tracy2016}
Tracy, P.~J., Kasper, J.~C., Raines, J.~M., {et~al.} 2016, Physical Review
  Letters, 255101, 255101

\bibitem[{Tracy {et~al.}(2015)Tracy, Kasper, Zurbuchen, Raines, Shearer, \&
  Gilbert}]{Tracy2015}
Tracy, P.~J., Kasper, J.~C., Zurbuchen, T.~H., {et~al.} 2015, The Astrophysical
  Journal, 812, 170, publisher: IOP Publishing

\bibitem[{Tu \& Marsch(1995)}]{Tu1995}
Tu, C.~Y. \& Marsch, E. 1995, Space Science Reviews, 73, 1, iSBN: 0038-6308

\bibitem[{Tu {et~al.}(1989)Tu, Marsch, \& Thieme}]{Tu1989}
Tu, C.-Y., Marsch, E., \& Thieme, K.~M. 1989, Journal of Geophysical Research,
  94, 11739

\bibitem[{Vauclair \& Charbonnel(1991)}]{Vauclair1991}
Vauclair, S. \& Charbonnel, C. 1991, in Challenges to {Theories} of the
  {Structure} of {Moderate}-{Mass} {Stars} (Berlin, Heidelberg: Springer Berlin
  Heidelberg), 37--41

\bibitem[{Verniero {et~al.}(2022)Verniero, Chandran, Larson, Paulson, Alterman,
  Badman, Bale, Bonnell, Bowen, de~Wit, Kasper, Klein, Lichko, Livi, McManus,
  Rahmati, Verscharen, Walters, \& Whittlesey}]{Verniero2022}
Verniero, J.~L., Chandran, B. D.~G., Larson, D.~E., {et~al.} 2022, The
  Astrophysical Journal, 924, 112, publisher: IOP Publishing

\bibitem[{Verniero {et~al.}(2020)Verniero, Larson, Livi, Rahmati, McManus,
  Pyakurel, Klein, Bowen, Bonnell, Alterman, Whittlesey, Malaspina, Bale,
  Kasper, Case, Goetz, Harvey, Korreck, MacDowall, Pulupa, Stevens, \&
  de~Wit}]{Verniero2020}
Verniero, J.~L., Larson, D.~E., Livi, R., {et~al.} 2020, The Astrophysical
  Journal Supplement Series, 248, 5, arXiv: 2004.03009 Publisher: IOP
  Publishing

\bibitem[{von Steiger {et~al.}(2000)von Steiger, Schwadron, Fisk, Geiss,
  Gloeckler, Hefti, Wilken, Wimmer-Schweingruber, \&
  Zurbuchen}]{vonSteiger2000}
von Steiger, R., Schwadron, N.~A., Fisk, L.~A., {et~al.} 2000, Journal of
  Geophysical Research: Space Physics, 105, 27217

\bibitem[{Weberg {et~al.}(2015)Weberg, Lepri, \& Zurbuchen}]{Weberg2015a}
Weberg, M., Lepri, S.~T., \& Zurbuchen, T.~H. 2015, Astrophysical Journal, 801,
  1

\bibitem[{Woodham {et~al.}(2018)Woodham, Wicks, Verscharen, \&
  Owen}]{Woodham2018}
Woodham, L.~D., Wicks, R.~T., Verscharen, D., \& Owen, C.~J. 2018, The
  Astrophysical Journal, 856, 49, arXiv: 1801.07344 Publisher: IOP Publishing

\bibitem[{Xu \& Borovsky(2015)}]{Xu2014}
Xu, F. \& Borovsky, J. 2015, Journal of Geophysical Research: Space Physics,
  120, 70

\bibitem[{Yardley {et~al.}(2024)Yardley, Brooks, D’Amicis, Owen, Long, Baker,
  Démoulin, Owens, Lockwood, Mihailescu, Coburn, Dewey, Müller, Suen,
  Ngampoopun, Louarn, Livi, Lepri, Fludra, Haberreiter, \&
  Schühle}]{Yardley2024}
Yardley, S.~L., Brooks, D.~H., D’Amicis, R., {et~al.} 2024, Nature Astronomy

\bibitem[{{Yogesh} {et~al.}(2021){Yogesh}, Chakrabarty, \&
  Srivastava}]{Yogesh:Ahe}
{Yogesh}, Chakrabarty, D., \& Srivastava, N. 2021, Monthly Notices of the Royal
  Astronomical Society: Letters, 503, L17, publisher: Oxford University Press

\bibitem[{{Yogesh} {et~al.}(2022){Yogesh}, Chakrabarty, \&
  Srivastava}]{Yogesh:CME}
{Yogesh}, Chakrabarty, D., \& Srivastava, N. 2022, Monthly Notices of the Royal
  Astronomical Society: Letters, 111, 106

\bibitem[{Zhao {et~al.}(2013)Zhao, Landi, \& Gibson}]{Zhao2013}
Zhao, L., Landi, E., \& Gibson, S.~E. 2013, The Astrophysical Journal, 773, 157

\bibitem[{Zhao {et~al.}(2022)Zhao, Landi, Lepri, \& Carpenter}]{Zhao2022}
Zhao, L., Landi, E., Lepri, S.~T., \& Carpenter, D. 2022, Universe, 8, 393

\bibitem[{Zhao {et~al.}(2017)Zhao, Landi, Lepri, Gilbert, Zurbuchen, Fisk, \&
  Raines}]{Zhao:InSituComposition:Sources}
Zhao, L., Landi, E., Lepri, S.~T., {et~al.} 2017, The Astrophysical Journal,
  846, 135, publisher: IOP Publishing

\bibitem[{Zurbuchen {et~al.}(2016)Zurbuchen, Weberg, Von~Steiger, Mewaldt,
  Lepri, \& Antiochos}]{Zurbuchen2016}
Zurbuchen, T.~H., Weberg, M., Von~Steiger, R., {et~al.} 2016, The Astrophysical
  Journal, 826, 10

\bibitem[{Ďurovcová {et~al.}(2019)Ďurovcová, Šafránková, \&
  Němeček}]{Durovcova2019}
Ďurovcová, T., Šafránková, J., \& Němeček, Z. 2019, Solar Physics, 294,
  97

\end{thebibliography}
\bibliographystyle{aa}

\end{document}